\documentclass[twocolumn,showpacs,amsmath,amssymb,10pt, aps]{revtex4-2}
\usepackage{graphicx,color}
\usepackage{subcaption}
\usepackage{soul}
\usepackage{bbm, bm}
\usepackage{color}
\usepackage{xcolor}
\usepackage{ulem} 

\usepackage[hidelinks, colorlinks=true, linkcolor=blue, urlcolor=blue, citecolor=blue]{hyperref}

\usepackage[T1]{fontenc}
\usepackage[cp1250]{inputenc}
\usepackage{amsfonts}
\usepackage{amssymb, amsbsy}
\usepackage{amsmath, bbm}
\usepackage{epstopdf}
\usepackage{physics}

\usepackage{caption}

\begin{document}
\title{
Optimization of two-photon absorption for three-level atom}

\author{Masood Valipour$^{1*}$, Gniewomir Sarbicki$^1$, Karolina S\l{}owik$^1$, Anita D\k{a}browska$^{2\dagger}$}

\affiliation{$^1$Institute of Physics, Faculty of Physics, Astronomy and Informatics,  Nicolaus Copernicus University, Grudzi\c{a}dzka 5/7, 87--100 Toru\'n, Poland,\\ $^2$Institute of Theoretical Physics and Astrophysics, Faculty of Mathematics, Physics and Informatics, University of Gda\'nsk, Wita Stwosza 57, 80--308 Gda\'nsk, Poland\\
$^*$masood.valipour@umk.pl, $^{\dagger}$anita.dabrowska@ug.edu.pl
}

\begin{abstract}
This work discusses the problem of optimal excitation of a three-level atom of ladder-configuration by light in the two-photon state and coherent light carrying an average of two photons. The applied atom-light interaction model is based on the Wigner-Weisskopf approximation. We characterize the properties of the optimal two-photon state that excites an atom perfectly, i.e. with probability equal to one. We find that the spectro-temporal shape of the optimal state of light is determined by the lifetimes of the atomic states, with the degree of photonic entanglement in the optimal state depending on the lifetime ratio. In consequence, two distinct interaction regimes can be identified in which the entanglement of the input state of light has a qualitatively different impact. 

As the optimal states may be challenging to prepare in general, we compare the results with those obtained for photon pairs of selected experimentally-relevant pulse shapes. As these shapes are optimized for maximal atomic excitation probability, the results can be interpreted in terms of the overlap between the optimal and investigated pulse shapes.

\end{abstract}

\maketitle

\section{Introduction and motivation}\label{sec:intro}
Two-photon absorption (TPA) is a nonlinear optical process where two photons are absorbed simultaneously by a material and excite a quantum system such as an atom or molecule \cite{Mollow1968}. In contrast to single-photon absorption, where one photon provides the necessary energy for excitation, TPA requires two photons, each supplying only part of the total energy needed for the transition. This process has a range of applications, including two-photon fluorescence microscopy, which enables high-resolution three-dimensional (3D) sample characterization \cite{Garcia2020}, photodynamic therapy \cite{Bhawalkar1997}, where excitation is confined to the focal point to minimize damage to surrounding tissues, and 3D optical data storage \cite{Strickler1991}. However, due to the inherently low cross-sections of two-photon events, high-power lasers are often required to achieve sufficient photon flux, potentially undermining the benefits of TPA and posing a risk of damaging sensitive samples.

A promising approach to addressing this problem exploits entangled photon pairs, which can be generated through spontaneous parametric down-conversion (SPDC) \cite{Kwiat1995}. These entangled photons behave as a single quantum entity, arriving at a molecule at correlated times, which may reduce significantly the photon flux required to achieve a comparable signal \cite{Javanainen1990,Tabakaev2021,Schlawin2024}. Entangled two-photon absorption (ETPA) has already been demonstrated in atomic vapours \cite{Georgiades1995,Dayan2004} and Rhodamine 6G molecules \cite{Tabakaev2021} in squeezed vacuum. Foundational theoretical work on TPA was established in early seminal studies \cite{GeaBanacloche1989,Javanainen1990}. More recently, perturbative frameworks have been developed to support experimental efforts and to optimize strategies by tailoring the spectro-temporal characteristics of entangled photon pairs \cite{schlawin2017,landes2021,Raymer2021,Carnio2021,Raymer2022, Giri2022}. The role of photon correlations in the context of vibronic selectivity in molecular entangled two-photon absorption was explained in Ref. \cite{Rodriguez-Camargo2024}.
An absorption enhancement strategy based on spectral phase flips was discussed in Ref. \cite{li2023}. A fully analytical stochastic framework to describe the TPA dynamics in three-level atomic systems has been introduced in \cite{DS24}. These advancements go hand in hand with the ongoing progress in engineering SPDC sources \cite{lutz2013,gajewski2016,pollmann2024,Serino2024} and refining detection schemes \cite{Panahiyan2023}.

In this work, the framework from Ref.\cite{DS24} is applied for exploration of the TPA dynamics in model three-level molecules. We demonstrate that the optimization of the spectro-temporal characteristics of the input light must take into account the molecular optical properties. We identify two distinct regimes of real and virtual states, defined through the ratio of lifetimes of the states contributing to the process, in which light entanglement has qualitatively different impact and role.

The starting point of our consideration is the analytical formula for the probability of two-photon absorption by a three-level atom interacting with light in a two-photon state, as determined in \cite{DS24}. We consider a two-photon state for a continuous-mode input field \cite{Loudon00}. This formula was derived using a stochastic approach in the input-output formalism \cite{GardinerZoller10,WisemanMilburn2010}, formulated within the Wigner-Weisskopf approximation \cite{Scully1997}. Note that the evolution of a quantum atomic system interacting with a wave packet of a definite number of photons is not governed by a single Gorini-Kossakowski-Sudarshan-Lindblad master equation \cite{GKS76,Lin76}, but by a set of coupled differential equations \cite{BCBC12, Dabrowska19}. For the photons in an entangled state, one obtains the set of infinitely many coupled equations. But, as shown in the paper \cite{DS24}, it can be solved analytically. 

We consider a bidirectional field prepared in a two-photon state. Our results can be straightforwardly generalized to the case of a multi-directional field carrying two photons or to take into account coupling with another input field in the vacuum state. We study the time-dependent probability of the excitation of the system. 
In the paper, we comprehensively analyze the properties of the optimal two-photon state, i.e., a state that gives a perfect excitation of a three-level system at some moment. Any deviations from this perfect match to the atom state suppress the maximum absorption probability. The optimal state becomes a reference point for understanding the conditions of maximized excitation of a three-level atom by light in other states.  We consider two-photon light states with uncorrelated photons and a state with entangled photons. We demonstrate the maximum excitation probabilities that can be achieved by optimally selecting pulse parameters, such as central frequencies, pulse widths, and the delay between the mean arrival times of the photons. We show how the parameters of light for optimal excitation are related to the properties of the atom. We also study the effect of pulse shape on the excitation of the atom. In addition, we compare the optimal excitation with two-photon light with the results for coherent pulses. We show how the effectiveness of the absorption process depends on the state of light and the features of the atom. We analyze the role of the entanglement of photons in the excitation process and show that the advantage of using the entangled state depends on the properties of the atom.
We identify two distinct interaction regimes, determined by the respective lifetimes of the intermediate and final states, in which entanglement of the input state of light has a qualitatively different impact. 

The structure of the paper is as follows. Section \ref{sec:model} contains a brief description of the model and the formula for the excitation of the final state of the atom. Section \ref{sec:optimal} is dedicated to presenting the two-photon state that optimally excites the atom. Section \ref{sec:unentangled} presents the effect of optimizing the excitation of the final state for the light in two-photon states with uncorrelated photons. Section \ref{sec:entangled} contains an analysis of the optimization of the final state excitation with entangled Gaussian two-photon states. 
Section \ref{sec:coherent} includes the optimization for the coherent pulses. Section \ref{sec:conclusions} provides a summary of the results. Appendices provide proofs and additional calculations. The effect of detuning the central frequencies of the pulses from the atomic transition frequencies on the excitation of the final state is analyzed in Appendix \ref{Appendix: Detunings}.

\section{Model}\label{sec:model}

\begin{figure}[h]\label{energy_scheme}
	\begin{center}
		\includegraphics[width=4cm]{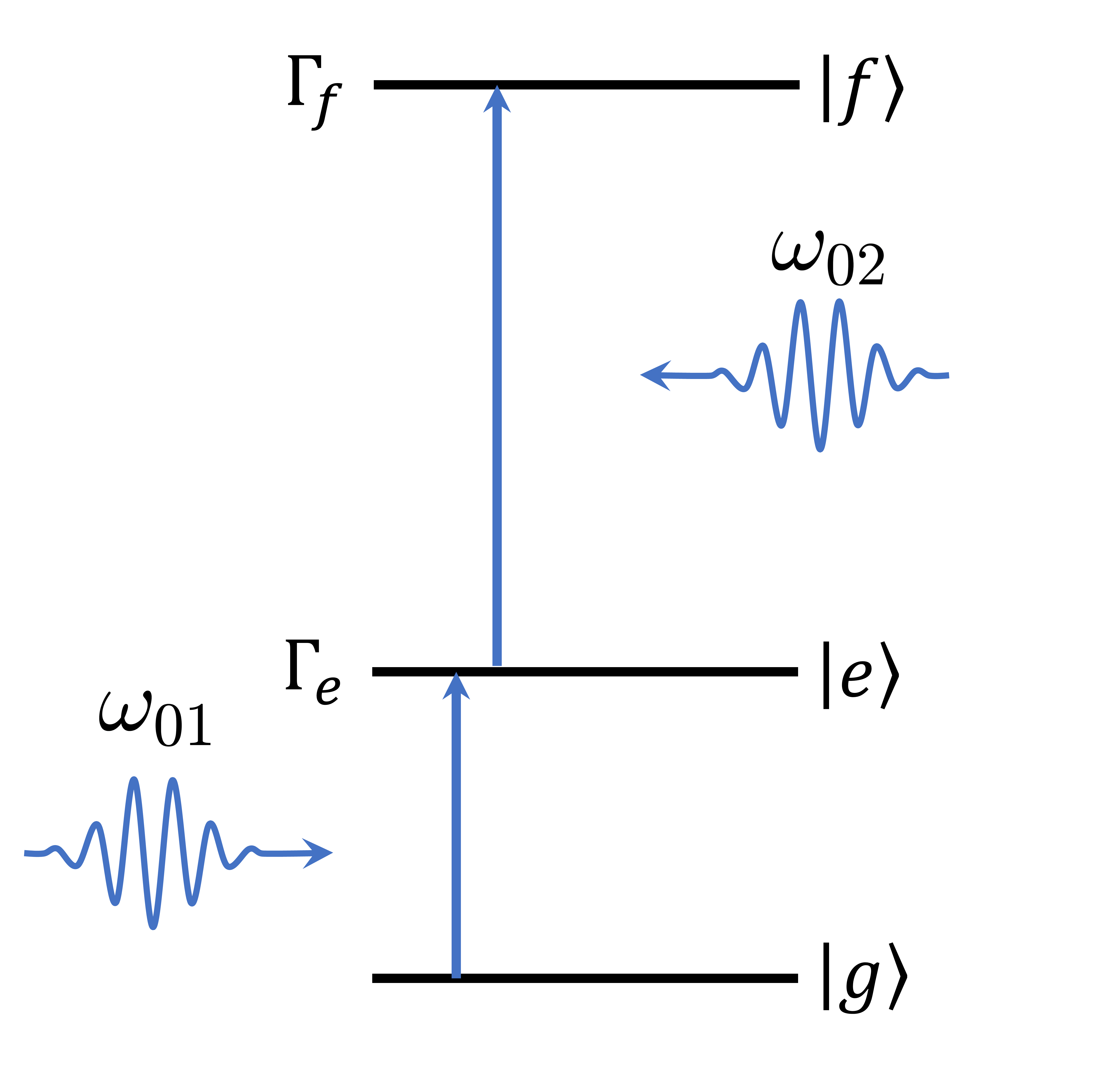}
		\caption{Energy scheme of three-level atom} \label{fig: energy_scheme}
	\end{center}
\end{figure}

We consider a propagating electromagnetic field interacting with a single three-level atom (Fig. \ref{fig: 
 energy_scheme}). We use the model of light-matter interaction  
 with the following assumptions:  the rotating-wave approximation, the flat coupling constant, and the extension of the lower limit of integration over frequency to minus infinity  \cite{Scully1997,Loudon00,GardinerZoller10,WisemanMilburn2010}. Thus, we assume that the bandwidth of the spectrum is much smaller than the central frequency of the pulse. We consider two independent field modes of different central frequencies, propagating in different directions. This field interacts with a model three-level atom in a ladder configuration with the states denoted by $\ket{g}$, $\ket{e}$, and $\ket{f}$. The transition frequencies are denoted as $\omega_{eg}$ for the lower transition and $\omega_{fe}$ for the upper one. We assume that each field mode couples to only one transition.

The evolution operator for the total system, consisting of the atom and the field, has the form \cite{GardinerZoller10}
 \begin{equation}\label{unitary}
 \hat{U}(t)=\overleftarrow{T}\exp\left[\int_{t_{0}}^{t}\!\left(-i\hat{H}+\sum_{k=1}^{2}\left(\hat{L}_k \hat{a}^{\dagger}_{k}(s)-\hat{L}^{\dagger}_{k}\hat{a}_{k}(s)\right)\!\right)\!ds\!\right],
 \end{equation}
 where $\overleftarrow{T}$ is the time-ordering operator and
 \begin{equation}
 \hat{a}_{k}(t)=\frac{1}{\sqrt{2\pi}}\int_{-\infty}^{+\infty}{ d}\omega\hat{a}_{k}(\omega){e}^{-{i}(\omega-\omega_{0k}) t},
 \end{equation}
where $\omega_{0k}$ represents the carrier frequency of the $k$-th input field. To simplify notation, we set $\hbar=1$. The unitary evolution is written in the interaction picture with respect to the free field dynamics. The time-dependent annihilation $\hat{a}_{k}(t)$ and creation $\hat{a}_{l}^{\dagger}(t^{\prime})$ operators satisfy the commutation relation
\begin{equation}
[\hat{a}_{k}(t),\hat{a}_{l}^{\dagger}(t^{\prime})]=\delta_{kl}\delta(t-t^{\prime}).
\end{equation}
The interaction between the atom and the bidirectional electromagnetic field is characterized by the two coupling operators:
 \begin{equation}\label{eq: coupling_operators}
 \hat{L}_1=\sqrt{\Gamma}_{e}\ketbra{g}{e},\;\;\hat{L}_2=\sqrt{\Gamma}_{f}\ketbra{e}{f},
 \end{equation} 
 where $\Gamma_e$ and $\Gamma_f$ are non-negative coupling constants, being the inverse lifetimes of the excited and final states due to the interaction with respective light modes. We work in the rotating frame defined by the unitary transformation
 \begin{equation}\label{eq: rotating_frame}
 \exp\left\{i\omega_{01}t\ketbra{g}{g}-i\omega_{02}t\ketbra{f}{f}\right\}.
 \end{equation}
 Thus, the Hamiltonian of the atomic system has the form
 \begin{equation}\label{eq: hamiltonian}
 \hat{H}=-\Delta_{1}\ketbra{g}{g}+ \Delta_{2}\ketbra{f}{f},
 \end{equation}
 where $\Delta_1$ and $\Delta_2$ are detunings defined respectively as $\Delta_1=\omega_{eg}-\omega_{01}$ and  $\Delta_2=\omega_{fe}-\omega_{02}$. By $t_0$, we denote the moment at which the systems start to interact.
 
We assume that the bidirectional field is prepared in the two-photon state vector  
\begin{equation}\label{eq: tps}
\ket{2_{\Psi}}= \int_{t_0}^{+\infty}dt_2\int_{t_0}^{+\infty}dt_1\Psi(t_2,t_1)\hat{a}^{\dagger}_2(t_2)\hat{a}^{\dagger}_{1}(t_1)
\ket{0,0},
\end{equation}
where $\Psi(t_2,t_1)$ is a two-dimensional complex function that satisfies the normalization condition
\begin{equation}
\int_{t_{0}}^{+\infty}dt_{2}\int_{t_{0}}^{+\infty}dt_1|\Psi(t_2,t_1)|^2=1.
\end{equation}
Let us recall that for an arbitrary two-photon state of the form (\ref{eq: tps}), there exists the Schmidt decomposition \cite{Parker00, Law2000}

\begin{equation}\label{eq: sd}
\Psi(t_2,t_1)=\sum_{n=0}^{+\infty}u_{n}\phi_{n}(t_2)\xi_{n}(t_1),
\end{equation}
where $\{\xi_{n}\}_{n=0}^{+\infty}$, $\{\phi_{n}\}_{n=0}^{+\infty}$ are two sets of orthogonal functions: 
\begin{equation}
\int_{t_0}^{+\infty}ds\xi_{n}(s)\xi_{m}^{\ast}(s)=\delta_{nm},
\end{equation}
\begin{equation}
\int_{t_0}^{+\infty}ds\phi_{n}(s)\phi_{m}^{\ast}(s)=\delta_{nm},
\end{equation}
and 
\begin{equation}\label{eq: norm_u}
\sum_{n=0}^{+\infty}u_{n}^2=1.
\end{equation}
It is assumed here that the phases are included into the definitions of the basis states, so coefficients in the Schmidt decomposition are real. 
The measure of entanglement of the two photons for a pure two-photon state is given by the von Neumann  entropy 
\begin{equation}\label{eq: entropy}
S= - \sum_{n=0}^{+\infty}u_{n}^2\log_{2}u_n^2.
\end{equation}
For unentangled photons, the amplitude of the state is given as a product  
\begin{equation}\label{eq: tps_uncorr}
\Psi(t_2,t_1)= \phi(t_2)\xi(t_1),
\end{equation}
where $\xi, \phi \in \mathbb{C}$ are temporal profiles of the photons that satisfy the normalization condition  
\begin{equation}
\int_{t_{0}}^{+\infty}ds|\xi(s)|^2=\int_{t_{0}}^{+\infty}ds|\phi(s)|^2=1.
\end{equation}
We assume that there is no initial correlation between the field and the atom and the state of the total system is given as   
 \begin{equation}
 \ket{2_{\Psi}}\otimes \ket{g}.
 \end{equation}
The evolution of a quantum system interacting with two-photon light was described using a stochastic approach and quantum trajectories in the paper \cite{DS24}. The analytical formulas determined in \cite{DS24} for the quantum trajectories allowed to find the formula for the probability of the excitation of the state $\ket{f}$ at time $t$ by the bidirectional light prepared in the state (\ref{eq: tps}),
\begin{widetext}
\begin{equation} \label{eq: Pf}
P_{f}(t)= \Gamma_f\Gamma_e \left|\int_{t_0}^{t}dt_2\exp\left[-\left(i\Delta_{2}+\frac{\Gamma_f}{2}\right)(t-t_2)\right]\int_{t_0}^{t_2}dt_1\exp\left[i\Delta_1 t_1-\frac{\Gamma_e}{2}(t_2-t_1)\right]\Psi(t_2,t_1)\right|^2,
\end{equation}
which can be rewritten to the form
\begin{equation} \label{eq: Pf_2}
P_{f}(t)= \Gamma_f\Gamma_e e^{-\Gamma_ft}\left|\int_{t_0}^{t}dt_2e^{\left(i\Delta_{2}+\frac{1}{2}(\Gamma_f-\Gamma_e)\right)t_2}\int_{t_0}^{t_2}dt_1e^{\left(i\Delta_1+\frac{\Gamma_e}{2} \right)t_1}\Psi(t_2,t_1)\right|^2.
\end{equation}
\end{widetext}
The latter formula is more compact; however, the form of eq.~(\ref{eq: Pf}) shows the sequential character of the absorption, where the photon in mode 1 excites the atomic system from the ground state $\ket{g}$ to the excited state $\ket{e}$ at time $t_1$, and then the photon in mode 2 is absorbed and induces a transition from $\ket{e}$ to the final state $\ket{f}$ at time $t_2 > t_1$. Note that losses induced by the coupling of the atomic system to its photonic environment in the vacuum can be easily incorporated by adding respective decaying exponential factors to the formula above. One can check that for any time $t\geq t_{0}$ the probability $P_{f}(t)$ is less than or equal to one. For the proof of it see Appendix \ref{Appendix: Proof}.  The properties that $P_{f}(t_0)=0$ and $\lim_{t\to \infty}P_{f}(t)=0$ reflect the fact that the atom starts in $\ket{g}$ and eventually relaxes to the ground state; thus, the excitation time of the atom by a pulse is finite.

The main goal of this work is to characterize how this probability can be maximized for different states of light depending on the characteristics of the atomic systems. In the next section, we discuss the state of light optimal for this purpose. However, this optimal state can be experimentally challenging to generate. Here, it will serve as the reference state. The following sections focus on experimentally feasible states, with Gaussian and exponential temporal shapes. In each case, we find qualitatively different results depending on atomic properties, covering the regime of similar excited and final state lifetimes, as well as the regime of a virtual excited state defined by the relation $\Gamma_e\gg \Gamma_f$.

\section{Optimal excitation}\label{sec:optimal}

\begin{figure*}[t!hb]
\includegraphics[width=15cm]{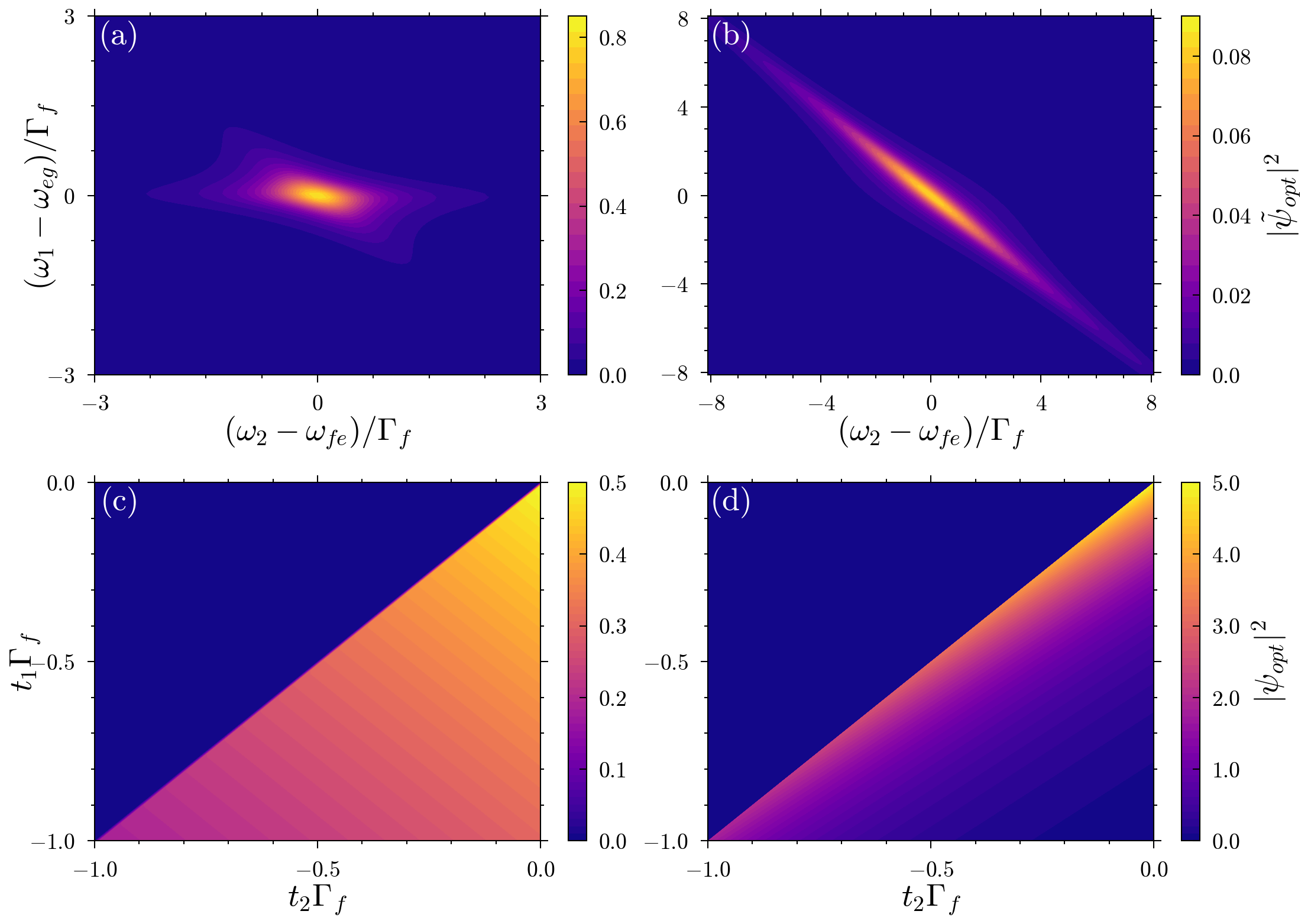}
\caption{Probability density functions in the frequency and time domains of the optimal two-photon state $\ket{2_{opt}}$ for the ratios:  $\Gamma_e/\Gamma_f = 0.5$ (plots (a) and (c)) and $\Gamma_e/\Gamma_f = 5$ (plots (b) and (d)).
} 
\label{fig: optimal}
\end{figure*}

In Ref.~\cite{DS24}, a proof was given that the maximal achievable value of the probability $P_{f}(t^\star)$ at time $t^\star > t_{0}$ is equal to
\begin{equation}\label{eq: P-max}
P^{\rm max}_{f}(t^\star) = 1-\frac{1}{\Gamma_f-\Gamma_{e}}\left(\Gamma_{f}e^{-\Gamma_{e}(t^\star-t_{0})} - \Gamma_{e}e^{-\Gamma_{f}(t^\star-t_{0}) }\right).
\end{equation}
In the case of $\Gamma_f = \Gamma_e=\Gamma$, the above formula 
(\ref{eq: P-max}) reduces to
        \begin{equation}
         P^{\rm max}_{f}(t^\star) =  1 - e^{-\Gamma\left(t^\star-t_0\right)} \left(1 + \Gamma \left(t^\star-t_0\right)\right).
        \end{equation}
We distinguish here the moment $t^\star$ at which the excitation probability is maximized. 
The probability of two-photon absorption reaches its maximum value on resonance, i.e., for $\Delta_1=\Delta_2=0$, and for the two-photon state of the temporal amplitude
\begin{equation}\label{eq: Phi-max}
 \Psi_{opt}(t_2,t_1)=\frac{1}{\sqrt{\mathcal{N}}} e^{\frac{1}{2}(\Gamma_f-\Gamma_e)t_2+\frac{1}{2}\Gamma_{e}t_{1}} \chi_{t_0<t_1<t_2<t^\star},  
\end{equation}
where $\mathcal{N}$ is the normalization factor having the form
\begin{align}\label{eq: normopt}
\mathcal{N}=&\frac{e^{\Gamma_f t^{\star}}}{\Gamma_{e}\Gamma_{f}}\left[1-e^{-\Gamma_f (t^{\star}-t_0)}\right.\nonumber\\&\left.+\frac{\Gamma_{f}}{\Gamma_{e}-\Gamma_{f}}\left(e^{-\Gamma_{e}(t^{\star}-t_0)}-e^{-\Gamma_{f}(t^{\star}-t_0)}\right)\right],
\end{align}
where $\chi_C$ is the characteristic function equal to 1 if the condition $C$ is fulfiled and 0 otherwise. 
Note that this pulse starts at $t_0$ and ends at time $t^\star$, for which the probability is maximized. Clearly, the photons in this state appear in a fixed order, i.e., first comes a photon exciting the transition $\ket{g}\rightarrow \ket{e}$, followed by a photon that excites the transition $\ket{e}\rightarrow \ket{f}$. Let us notice that the amplitude (\ref{eq: Phi-max}) can not be written as a product of two single-photon profiles meaning that we deal here with the entangled photons. Taking $t_{0}\to -\infty$, we obtain the optimal two-photon state of light of the form
\begin{widetext}
\begin{equation}\label{eq: optstate1}
\ket{2_{opt}}=\sqrt{\Gamma_e\Gamma_f}e^{-\frac{\Gamma_f t^{\star}}{2}}\int_{-\infty}^{t^{\star}}dt_2\int_{-\infty}^{t_2}dt_1e^{\frac{1}{2}(\Gamma_f-\Gamma_{e})t_2+\frac{1}{2}\Gamma_e t_1}\hat{a}^{\dagger}_{2}(t_2)\hat{a}^{\dagger}_{1}(t_1)\ket{0,0},
\end{equation}
\end{widetext}
which gives a perfect excitation, i.e., the probability $P_{f}(t^{\star})=1$. Therefore, we can conclude that for any time $t$ there exists the two-photon state of light for which the probability of two-photon absorption at time $t$ is equal to one.  Interestingly, the state (\ref{eq: optstate1}) is 
a time reversal of the two-photon state of light emitted spontaneously by the three-level atom in the ladder configuration prepared initially in the state $\ket{f}$. The state of light emitted in spontaneous emission by the three-level atom in the ladder configuration can be found in Ref. \cite{Scully1997}. An analogy can be seen here with the known result for the optimal excitation of a two-level atom by a pulse rising exponentially in time that is a time-reversal of the light spontaneously emitted by the two-level atom \cite{Leuchs2009,Scarani11,Banacloche17}.

The mean time of residence of the atom in the state $\ket{f}$ is defined as
\begin{equation}
{\tau}_{r}=\int_{-\infty}^{+\infty}dtP_{f}(t).
\end{equation}
One can check then that for the optimal state (\ref{eq: optstate1}), we obtain \mbox{${\tau}_{r}=2/\Gamma_f$}.
This is the maximum possible mean time of residence in the state $\ket{f}$ for the atom driven by the light in the two-photon state (\ref{eq: tps}). 

In the frequency domain, 
the state (\ref{eq: optstate1}) has the amplitude
\begin{align}
\tilde{\Psi}_{opt}(\omega_2,\omega_1)\!=\!\frac{(2\pi)^{-1}\sqrt{\Gamma_e\Gamma_{f}}e^{i(\omega_1+\omega_2-\omega_{fg})t^{\star}}}{\big[ i\left(\omega_1-\omega_{eg}\right)+\frac{\Gamma_e}{2}\big]\big[i\left(\omega_1+\omega_{2}-\omega_{fg}\right)+\frac{\Gamma_f}{2}\big]}.
\end{align}
Note that this amplitude is a function of $t^\star$, being the moment of optimal excitation. However, the biphoton probability density function $|\tilde{\Psi}_{opt}(\omega_2,\omega_1)|^2$ is independent of that moment. 
The formula shows the role of single and double resonance in the excitation of the atom from the state $\ket{g}$ to the state $\ket{f}$. Their respective impact depends on the atomic properties. Fig.~\ref{fig: optimal}(a) shows that in the regime of similar excited- and final-state lifetimes, it is important to tune light near both single- and two-photon resonance. Contrary, in the virtual-state regime, i.e., for $\Gamma_e\gg\Gamma_f$, the two-photon resonance plays the dominant role in the probability distribution. This effect is already apparent, e.g., for $\Gamma_e/\Gamma_f=5$, see Fig.~\ref{fig: optimal}(b).  We will use in this paper as a reference a scale determined by the lifetime $\Gamma_f^{-1}$ of the state $\ket{f}$.  The distributions over time correspond to $t^{\star}=0$.

 The marginal distribution for the frequency of the first photon, i.e., the photon tuned to the transition $\ket{g}\rightarrow \ket{e}$, is given by the Lorentzian function,
\begin{equation}\label{eq:p_w1}
p_1(\omega_1)= \frac{\Gamma_e}{2\pi\left[ \left(\omega_1-\omega_{eg}\right)^2+\frac{\Gamma_e^2}{4}\right]},
\end{equation}
while the marginal distribution for the frequency of the second photon, i.e., the photon tuned to the transition $\ket{e}\rightarrow \ket{f}$, by 
\begin{equation}\label{eq:p_w2}
p_2(\omega_2)= \frac{\Gamma_e+\Gamma_f}{2\pi\left[ \left(\omega_2-\omega_{fe}\right)^2+\frac{(\Gamma_e+\Gamma_f)^2}{4}\right]}. 
\end{equation}
This is a Lorentzian distribution with the broadening being a sum of the broadenings of states $\ket{e}$ and $\ket{f}$. One can check that the conditional probability density function of the frequency $\omega_2$ of the second photon given the frequency $\omega_1$ of the first photon, reads as 
\begin{equation}
p(\omega_2|\omega_1)= \frac{\Gamma_f}{2\pi \left[\left(\omega_{2}-(\omega_{fg}-\omega_1)\right)^2+\frac{\Gamma_f^2}{4}\right]}.
\end{equation}
Note that the Lorentzian spectral distributions for the two photons in Eqs.~(\ref{eq:p_w1}) and (\ref{eq:p_w2}) become similar in the regime of $\Gamma_e\gg\Gamma_f$, i.e., 
\begin{equation}
p_2(\omega_2)\xrightarrow[]{{\Gamma_{e}}\gg{\Gamma_f}} \frac{\Gamma_e}{2\pi\left[ \left(\omega_2-\omega_{fe}\right)^2+\frac{\Gamma_{e}^2}{4}\right]}.
\end{equation}
As shown in Sec. IV, the similarity of the marginal spectral width of the photons in the optimized states of light in this regime is indeed observed.

Let us notice that for the optimal state (\ref{eq: optstate1}), the joint probability density of the times of arrival of the photons at the position of the atomic system is defined by the expression
\begin{equation}
p(t_2,t_1)= \Gamma_f\Gamma_e e^{-\Gamma_f(t^{\star}-t_2)-\Gamma_e(t_2-t_1)},
\end{equation}
where $t_1<t_2\leq t^{\star}$.
Hence, one can get the probability density of the time of arrival of the first photon, 
\begin{eqnarray}\label{eq: p_1}
p_1(t_1) &=& \int_{t_1}^{t^{\star}}dt_2p(t_2,t_1)\nonumber\\
&=&\frac{\Gamma_e\Gamma_f}{\Gamma_f-\Gamma_e}\left(e^{-\Gamma_{e}(t^{\star}-t_1)}-e^{-\Gamma_f(t^{\star}-t_1)}\right) 
\end{eqnarray}
and  the second photon,
\begin{equation}
p_2(t_2)= \int_{-\infty}^{t_2}dt_1p(t_2,t_1)= \Gamma_fe^{-\Gamma_f(t^{\star}-t_2)}.
\end{equation}
Therefore, the expected times of appearance of photons at the atomic position, defined by 
\begin{equation}\tau_{i}=\int_{-\infty}^{t^{\star}}dt_it_ip_i(t_i),
\end{equation}
are given, respectively, by
\begin{equation}
\tau_1= t^{\star}-\frac{1}{\Gamma_e}-\frac{1}{\Gamma_f},\;\;\;
\tau_2 = t^{\star}-\frac{1}{\Gamma_f}.
\end{equation}
Thus their difference is equal to $1/\Gamma_e$. Clearly, this also represents the difference between the expected times of spontaneous emissions of the photons for our atomic system. Note that dealing with the state for which the excitation of the state $\ket{f}$ is perfect ($P_{f}(t^{\star})=1$), these quantities can also be identified as the mean moments of the photon absorption events.

The optimal state (\ref{eq: optstate1}) is an entangled state of two photons. The degree of entanglement of these photons can be represented as a function of the $\Gamma_e/\Gamma_f$ ratio, see Fig.~\ref{fig: shannon} Note that the larger the $\Gamma_e/\Gamma_f$ ratio, the more entangled the optimal state is.

To better understand the two different regimes of optimal excitation of the three-level system, it is useful to write down the probability density of the time of arrival of the first photon (\ref{eq: p_1}) in the form of
\begin{equation}
p_1(t_1) = \frac{\Gamma_ee^{-\Gamma_{e}(t^{\star}-t_1)}}{1-\frac{\Gamma_e}{\Gamma_f}}+\frac{\Gamma_fe^{-\Gamma_{f}(t^{\star}-t_1)}}{1-\frac{\Gamma_f}{\Gamma_e}}.
\end{equation}
Hence, we obtain two limiting expressions:
\begin{displaymath}
p_1(t_1)\xrightarrow[]{{\Gamma_{e}}\ll{\Gamma_f}} \Gamma_e e^{-\Gamma_{e}(t^{\star}-t_1)},  
\end{displaymath}
\begin{displaymath}
p_1(t_1)\xrightarrow[]{{\Gamma_{e}}\gg{\Gamma_f}} \Gamma_f e^{-\Gamma_{f}(t^{\star}-t_1)}.  
\end{displaymath}
Thus, if $\Gamma_e\ll \Gamma_f $, meaning that the lifetime of the excited state is much longer than that of the final state, then the atom is optimally excited with weakly correlated, almost independent photons. 
If $\Gamma_e\gg\Gamma_f$, then the photons of the optimal state are strongly correlated and the marginal distributions of photon arrival times approach each other.
This is an important conclusion of this work.

Note that the probability density function
of the sum of the frequencies for the light in the optimal state  has the form 
\begin{equation}
p_{+}(\omega_1+\omega_2) = \frac{\Gamma_{f}}{2\pi \left[\left(\omega_1+\omega_{2}-\omega_{fg}\right)^2+\frac{\Gamma_f^2}{4}\right]}. 
\end{equation}
while the density function of the difference of the frequencies is given by
\begin{equation}
p_{-}(\omega_2-\omega_1) = \frac{\Gamma_{f}+2\Gamma_e}{2\pi \left[\left(\omega_2-\omega_{1}-(\omega_{fe}-\omega_{eg})\right)^2+\frac{(\Gamma_f+2\Gamma_e)^2}{4}\right]}. 
\end{equation}
We will use these distributions as reference points later in the article. 

Above, we have discussed the optimal state that ensures perfect excitation of the three-level system at a selected moment $t^\star$. 
We proceed to investigate the two-photon absorption probability for other families of states standard in experimental realizations. We consider unentangled photons with Gaussian and exponential profiles 
and entangled photons with Gaussian distributions of sum and difference frequencies. Our focus is the value of the probability of atomic excitation maximized over times
\begin{equation}
\underset{t\in \mathbbm{R}}{\rm{max}} P_{f}(t).
\end{equation}
and the parameters of the light state. As shown in Appendix \ref{Appendix: Inner_Product}, the two-photon absorption probability can be interpreted as the inner product of the two functions defined respectively by the temporal amplitudes of the optimal state (\ref{eq: optstate1}) and of the two-photon state of the light driving the atom. The best excitation conditions are obtained at the resonances, i.e., $\Delta_1=0$, $\Delta_2=0$. The effect of the resonance detunings on atomic excitation is described in Appendix \ref{Appendix: Detunings}. 

\section{Results for two-photon states of unentangled photons}\label{sec:unentangled}

\begin{figure}[t]
\captionsetup{justification=justified}
\centering
\includegraphics[width=7cm]{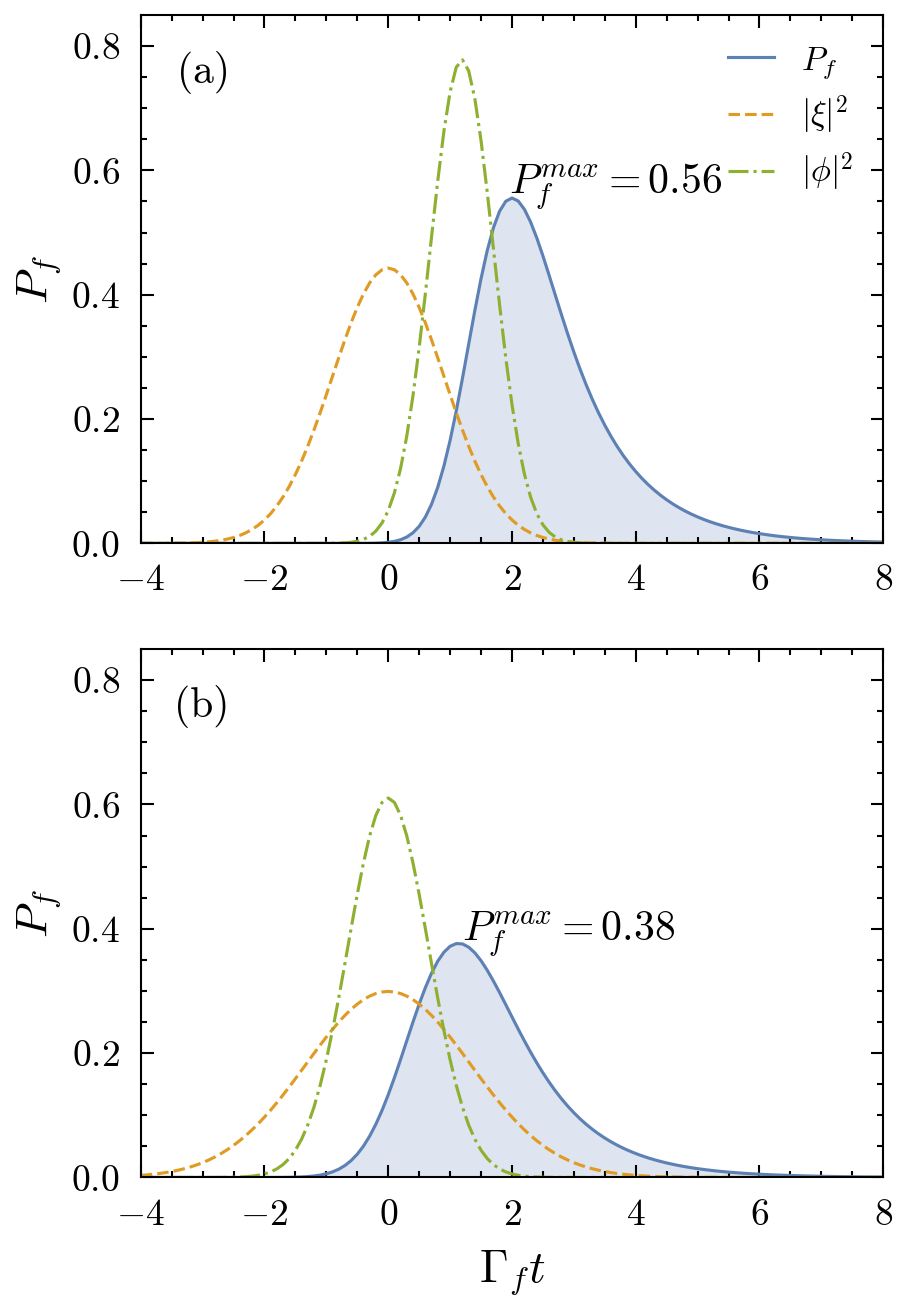}
\caption{Probability $P_{f}(t)$ as a function of $\Gamma_{f} t$ for the input two-photon state of unentangled photons with the Gaussian shapes for the ratio $\Gamma_e/ \Gamma_f = 1$ and the optimized parameters of light with (a) and without (b) photon delay allowed. 
The values of parameters are:
(a): $\Omega_1=0.75\Gamma_f$, $\Omega_2=1.53\Gamma_f$, $\mu=1.19/\Gamma_f$
and 
(b): $\Omega_1=1.11\Gamma_f$, $\Omega_2=1.95\Gamma_f$.
The dashed lines represent the input wave shapes with respect to $\Gamma_{f} t$.}

\label{fig:uncorrelated_Gaussian_Pf}
\end{figure}

\begin{figure}[t]
		\includegraphics[width=7cm]{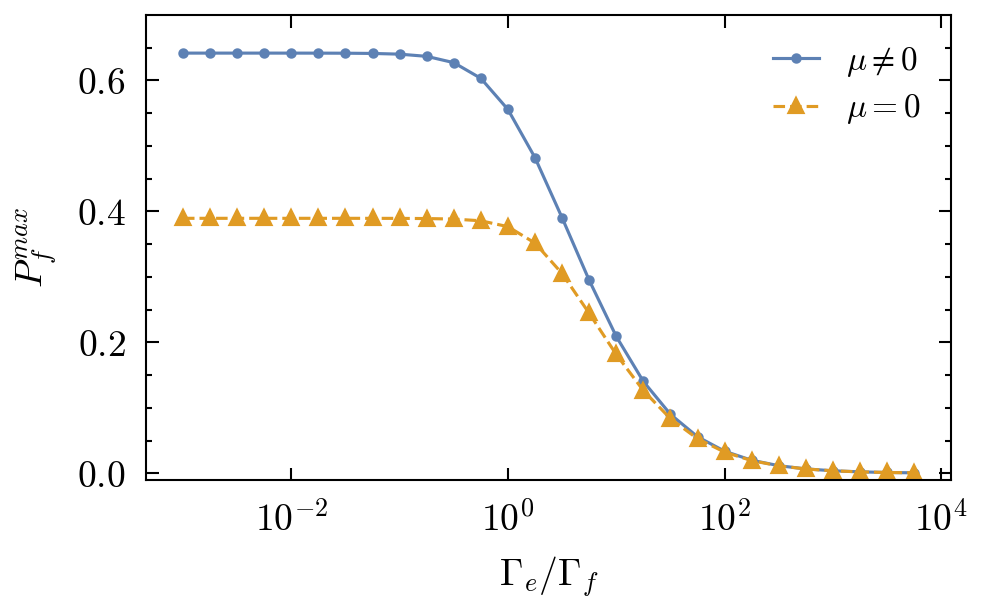}
		\caption{Maximum probability $P^{max}_{f}$ for the input two-photon state of unentangled photons with the Gaussian shapes, as a function of the ratio $\Gamma_{e}/\Gamma_{f}$ with the optimized bandwidth, and with and without optimized time delay. }
 \label{fig: Uncorr_Gaussian_GammaRatio}
\end{figure}

\begin{figure}[h]
\includegraphics[width=6.9cm]{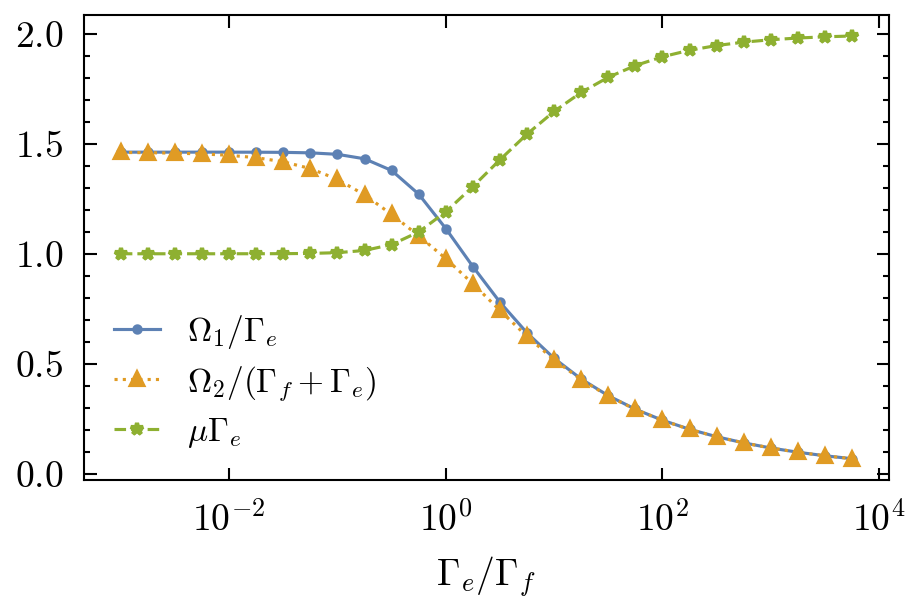}
\caption{Terms $\Omega_{1}/\Gamma_f$, $\Omega_{2}/(\Gamma_f+\Gamma_e)$, and $\mu\Gamma_e$ for optimal excitation of the atom by unentangled photons of the Gaussian shapes with different mean arrival times as a function of the $\Gamma_e/\Gamma_f$ ratio.}\label{fig: parameters_uncorr}
\end{figure}

In this section, we study the problem of optimal conditions for an excitation of the atom by two independent single-photon pulses. We start with a detailed description of the case when the atom is driven by the pulses of Gaussian shapes: 
\begin{equation}
\xi(t)= \left(\frac{\Omega_{1}^2}{2\pi}\right)^{1/4}\exp\left(-\frac{\Omega_1^2 t^2}{4}\right),
\end{equation}
\begin{equation}
\phi(t)= \left(\frac{\Omega_{2}^2}{2\pi}\right)^{1/4}\exp\left(-\frac{\Omega_2^2 (t-\mu)^2}{4}\right),
\end{equation}
where $\Omega_1,\Omega_2>0$ are the spectral widths of the Gaussian profiles, and $\mu$ is the time delay between the maxima of the two pulses. Note that this quantity represents a quantum-mechanical average of the arrival time difference of the two photons. 
These shapes were optimized for a maximized probability (\ref{eq: Pf}), where we have not predefined the time $t^\star$ at which the maximal value is obtained. Here we present the results for the resonances: $\Delta_1=\Delta_2=0$.
We show the effect of the optimal choice of the pulses' bandwidths and the time delay of the second photon for the different values of the ratio $\Gamma_e/\Gamma_f$.
As an example, we present the results for equal excited- and final-state lifetimes, i.e. for $\Gamma_e=\Gamma_f$. Fig.~\ref{fig:uncorrelated_Gaussian_Pf}(a) shows the squared optimized profiles and the time-dependent probability of excitation of the final state for this case.  Note that the time delay $\mu$ was not restricted to nonnegative values.
In the optimal excitation conditions, the second photon is systematically narrower in the time domain: we observe this behaviour for different $\Gamma_e/\Gamma_f$ ratios. This can be explained by the finite lifetime of the excited state. Its expected arrival time is also delayed with respect to the one of the first photon, so that the second photon arrives at the atom when the probability of the excitation to the state $|e\rangle$ is already significant. We observe that the value of the optimal time delay is of the order of $\Gamma_e^{-1}$. The resulting final-state excitation probability has a maximum slightly after the expected arrival times of both photons, and decays smoothly afterwards. In Fig.~\ref{fig:uncorrelated_Gaussian_Pf}(b), we present the same quantities in the case of a vanishing time delay (we fixed here $\mu=0$), demonstrating that through a proper selection of the time delay $\mu$, it is possible to enhance the two-photon absorption probability significantly. The final-state excitation probabilities for these two scenarios are compared as a function of the ratio $\Gamma_e/\Gamma_f$ in Fig.~\ref{fig: Uncorr_Gaussian_GammaRatio}.  Note that in the case of $\Gamma_e\ll\Gamma_f$, the maximal probability approaches the value of $0.64$, being the squared maximal probability of excitation of a two-level system by a single photon of a Gaussian shape \cite{Scarani11,Banacloche17}.
This is achieved in the optimal case with the properly adjusted time delay between the photon wavepackets. This result can be intuitively understood as a sequence of two independent excitations of two-level systems $|g\rangle\leftrightarrow |e\rangle$, and $|e\rangle\leftrightarrow |f\rangle$.
As the $\Gamma_e/\Gamma_f$ ratio increases, the maximal final-state excitation probability is reduced, and the time delay between the photons leads to relatively smaller improvement of the probability. 
Finally, in the virtual-state limit, the final-state excitation with uncorrelated photons becomes inefficient. This observation is in line with the previous conclusion that the optimal state, given by (\ref{eq: optstate1}), becomes more entangled in the virtual-state regime. In both cases, with optimization of the arrival delay of the second photon and with equal photon arrival times, the parameters $\Omega_2/\Gamma_f$ and $\Omega_1/\Gamma_f$ increase as the $\Gamma_e/\Gamma_f$ ratio increases and their ratio $\Omega_2/\Omega_1$ tends to one. The optimal choice of light parameters for excitation of the atomic system is displayed for different $\Gamma_e/\Gamma_f$ ratios in Fig. \ref{fig: parameters_uncorr}. The spectral widths of the pulses are shown in relation to the widths at half-maximum of Lorentz distributions of the marginal spectral distribution for the first and second photon of the optimal state, equal to $\Gamma_e$ and $\Gamma_e+\Gamma_f$, respectively. Note that in the limit of $\Gamma_e\ll \Gamma_f$, the terms $\Omega_1/\Gamma_e$ and $\Omega_2/(\Gamma_e+\Gamma_f)$ tend to the value of $1.46$ as it is in the case when a two-level atom is optimally excited by a single-photon light 
\cite{Scarani11,Banacloche17}. The difference between the average photon arrival times is presented in relation to the lifetime of the middle state. It is seen that as $\Gamma_e \ll \Gamma_f$, then $\mu\Gamma_e$ approaches $1$, while for $\Gamma_e \gg \Gamma_f$, it approaches $2$. The sensitivity of the maximum excitation probability of the atom to the choice of the $\Omega_{1}$ and $\Omega_{2}$ parameters is discussed in Appendix \ref{Appendix: sensitivity}.

Results for rising and decaying exponential pulse shapes are described in detail in Appendix \ref{Appendix: Rising_Decaying_Shapes}. In the investigated range of parameters we confirm better performance of the rising over decaying exponential shapes. The impact of pulse shapes is less important in the virtual regime of $\Gamma_e\gg\Gamma_f$. For $\Gamma_e\ll\Gamma_f$, for the optimal choice of parameters the two-photon absorption dynamics boils down to a sequence of two single-photon absorption events. The achieved maximal TPA probability is a squared maximal single-photon absorption probability for each pulse shape.  
In particular, for the rising exponential pulses, the limiting case of the optimal shape is realized and the perfect excitation is achieved, i.e.,  $P_f^{max}=1$.

\section{Results for the two-photon state of entangled photons}\label{sec:entangled}

We consider in this section the two-photon state with the temporal amplitude of the form
\begin{equation}\label{eq: entanstate}
\Psi(t_2,t_1) = \sqrt{\frac{\Omega_{+}\Omega_{-}}{2\pi}}e^{-\frac{\Omega_{+}^2}{8}\left(t_2-\mu+t_1\right)^2-\frac{\Omega_{-}^2}{8}\left(t_2-\mu-t_1\right)^2},
\end{equation}
where $\Omega_{+},\Omega_{-}>0$. 
The width $\Omega_{+}$ is the spectral width of the laser pulse driving SPDC, while $\Omega_{-}$ is determined by phase matching \cite{Walmsley2001}.
Note that if $\Omega_{-}\neq \Omega_{+}$, we deal with an entangled state of the two photons. 
In the frequency domain, the considered state has the amplitude of the form
\begin{widetext}
    \begin{equation}
\tilde{\Psi}(\omega_2,\omega_1)=\sqrt{\frac{2}{\pi\Omega_{+}\Omega_{-}}}\exp \left\{-\frac{\left(\omega_2+\omega_1-(\omega_{02}+\omega_{01})\right)^2}{2\Omega_{+}^2}-\frac{\left(\omega_2-\omega_1-(\omega_{02}-\omega_{01})\right)^2}{2\Omega_{-}^2}+i(\omega_2-\omega_{02})\mu\right\}.
\end{equation}
\end{widetext}
One can check that for the light in the state (\ref{eq: entanstate}), the widths of the marginal spectral and temporal distributions for the two photons are the same. The variance value in the time domain is given by 
\begin{equation}
\sigma_{t}^2 = \frac{\Omega_+^2+\Omega_{-}^2}{2\Omega_{+}^2\Omega_{-}^2}
\end{equation} 
while in the frequency domain it is
\begin{equation}
\sigma_{\omega}^2= \frac{\Omega_{+}^2+\Omega_{-}^2}{8}.
\end{equation}
The probability density functions of the sum and the difference of the photon frequencies have the respective forms
\begin{equation}
p_{\pm}(\omega_1\pm\omega_2)= \frac{1}{\sqrt{\pi}\Omega_{\pm}}\exp(-\frac{\left(\omega_1\pm\omega_2-(\omega_{01}\pm\omega_{02})\right)^2}{\Omega_{\pm}^2}).
\end{equation}

\begin{figure}
\centering
\includegraphics[width=7cm]{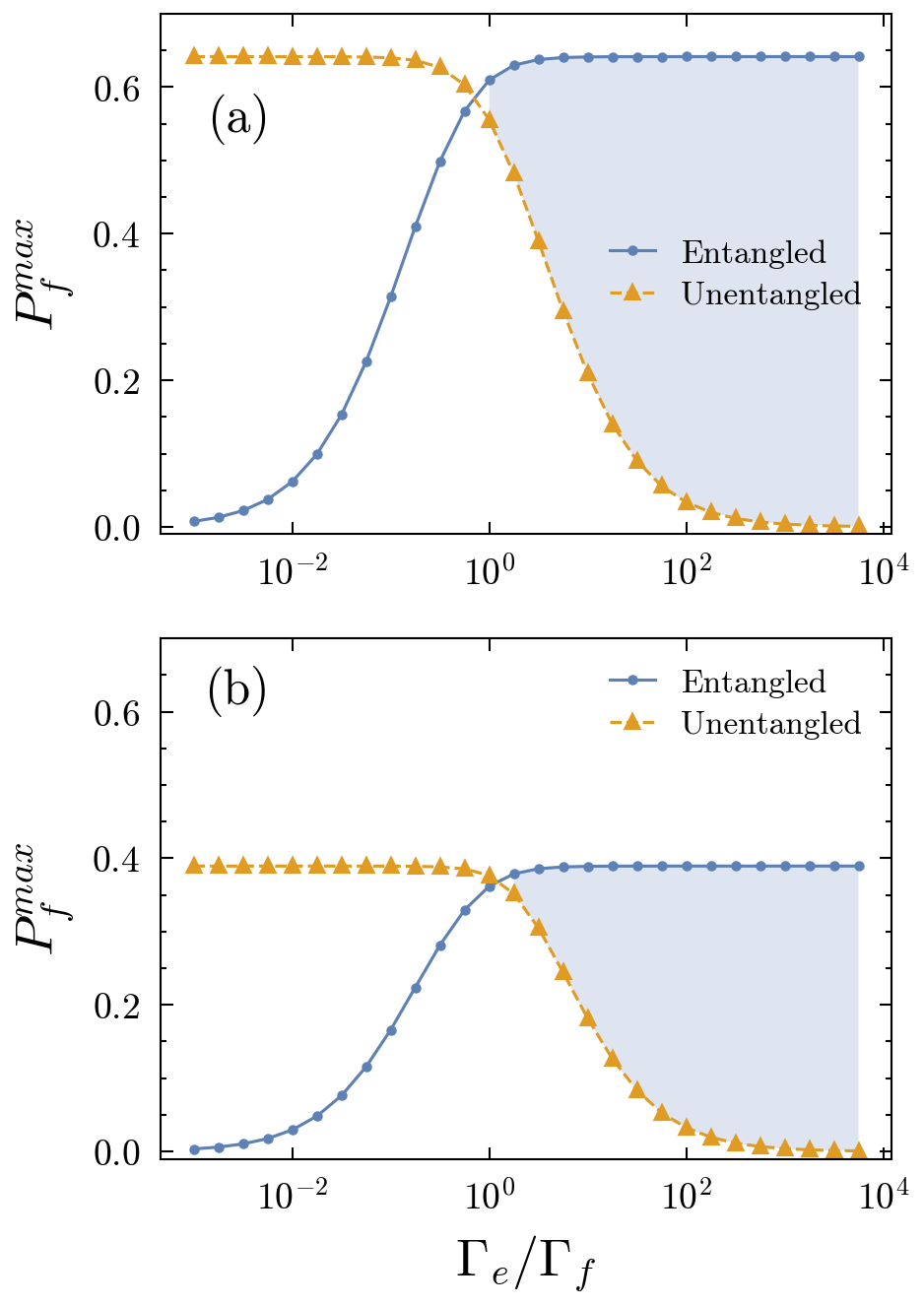}
\caption{Maximum probability $P^{max}_{f}$ as a function of the ratio $\Gamma_{e}/\Gamma_{f}$  for the input two-photon states for entangled and unentangled photons with the optimized bandwidth, and (a) with optimized (b) without time delay $\mu$. The shaded area shows the improvement due to the entanglement between photons.
} 
\label{fig:corr_vs_entangled}
\end{figure}

We optimize the state of the photon pair for a maximized final-state excitation probability in the atom. Here we present the results for the resonant case. Two scenarios are also compared here: one with the optimization of the average difference between the photon arrival times, and the other with identical average times. As before, the scenario with the nonzero time delay brings the benefit of an increased probability, as shown in Fig.~\ref{fig:corr_vs_entangled} as a function of the $\Gamma_e/\Gamma_f$ ratio. We verified that in this case, the value of the optimal delay is also of the order of $\Gamma_e^{-1}$. Contrary to the uncorrelated case, the final state is efficiently excited for a broad range of $\Gamma_e/\Gamma_f$ ratios, including the virtual-state regime. 
This can be understood, since as the excited-state lifetime decreases, the role of correlations of the photon-arrival times becomes more significant. For small $\Gamma_e/\Gamma_f$ ratios, the excitation scenario exploiting entangled pairs is sub-optimal and results in decreased final-state probabilities. 
Note that in this regime, the degree of entanglement of the optimal state (\ref{eq: optstate1}) tends to zero.  For comparable $\Gamma_e$ and $\Gamma_f$, there is little benefit from using entangled states. Interestingly, the values of the maximum achievable probabilities of excitation of the final state correspond to the maximum probabilities for the case of uncorrelated photons with Gaussian shapes.


\begin{figure}[t]
    \includegraphics[width=7cm]{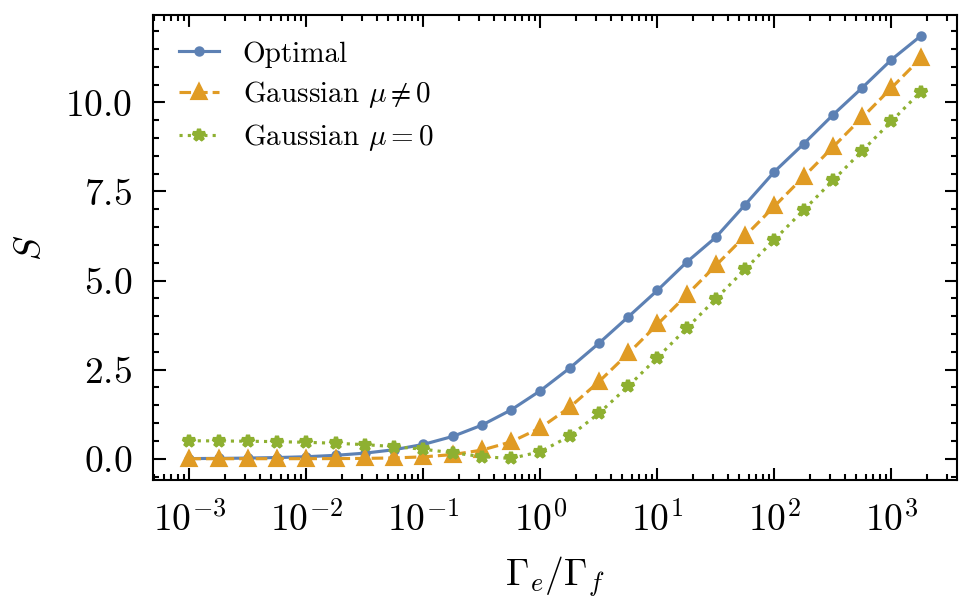}
		\caption{The entropy of entanglement $S$ as a function of the ratio $\Gamma_e/\Gamma_f$ for optimized light parameters, with and without optimized time delay between the photons. }
 \label{fig: shannon}
\end{figure}

We further investigate the role of entanglement using Schmidt decomposition. For the two-photon state (\ref{eq: entanstate}), we have
\begin{equation}
\Psi(t_2,t_1)= \sum_{n=0}^{+\infty}u_{n}\phi_{n}(t_2-\mu)\phi_{n}(t_1),
\end{equation}
where
\begin{equation}
u_{n}= 2\frac{\sqrt{\Omega_{+}\Omega_{-}}}{\Omega_{+}+\Omega_{-}}\left(\frac{\Omega_{-}-\Omega_{+}}{\Omega_{-}+\Omega_{+}}\right)^n
\end{equation}
and
\begin{equation}
    \phi_{n}(t)=\sqrt[4]{\frac{\Omega_{-}\Omega_{+}}{2\pi}}
\frac{1}{\sqrt{2^{n}n!}}e^{-\frac{1}{2}\left(\sqrt{\frac{\Omega_{+}\Omega_{-}}{2}}t\right)^2} H_{n}\left(\sqrt{\frac{\Omega_{+}\Omega_{-}}{2}}t\right). 
\end{equation}
Here, $H_{n}$ are the 
Hermite polynomials of degree $n$ defined by the Rodrigues formula 
\begin{equation}
H_{n}(x)=(-1)^{n}e^{x^2}\frac{d^n}{dx^n}e^{-x^2}
\end{equation}
for all $x\in \mathbbm{R}$ and $n=0,1,2,\ldots$.
One can check that the entanglement entropy (\ref{eq: entropy}) in this case can be expressed as 
\begin{equation}
S = -\log_{2}(1-y)-\frac{y}{1-y}\log_{2}y,
\end{equation}
where $y=(\Omega_{-}-\Omega_{+})^2/(\Omega_{-}+\Omega_{+})^2$. Fig.~\ref{fig: shannon} shows the Shannon entropy for the optimized entangled two-photon state with $\mu\neq 0$ and  $\mu=0$.
In the broad range of $\Gamma_e/\Gamma_f$ ratios where the photon correlation is beneficial, the optimal state with $\mu\neq 0$ has a higher degree of entanglement than the state for which $\mu=0$ is enforced. 
There, the degree of entanglement of the optimized state increases with the $\Gamma_e/\Gamma_f$ ratio.
On the contrary, for small $\Gamma_e/\Gamma_f$ ratios, the entanglement becomes detrimental, similarly to the scenario when the mean times of arrival of the photons are taken equal. In this latter case, we observe for $\Gamma_e/\Gamma_f <0.5$ a benefit from photon anticorrelation in the time domain. 

\begin{figure}
\includegraphics[width=8.5cm]{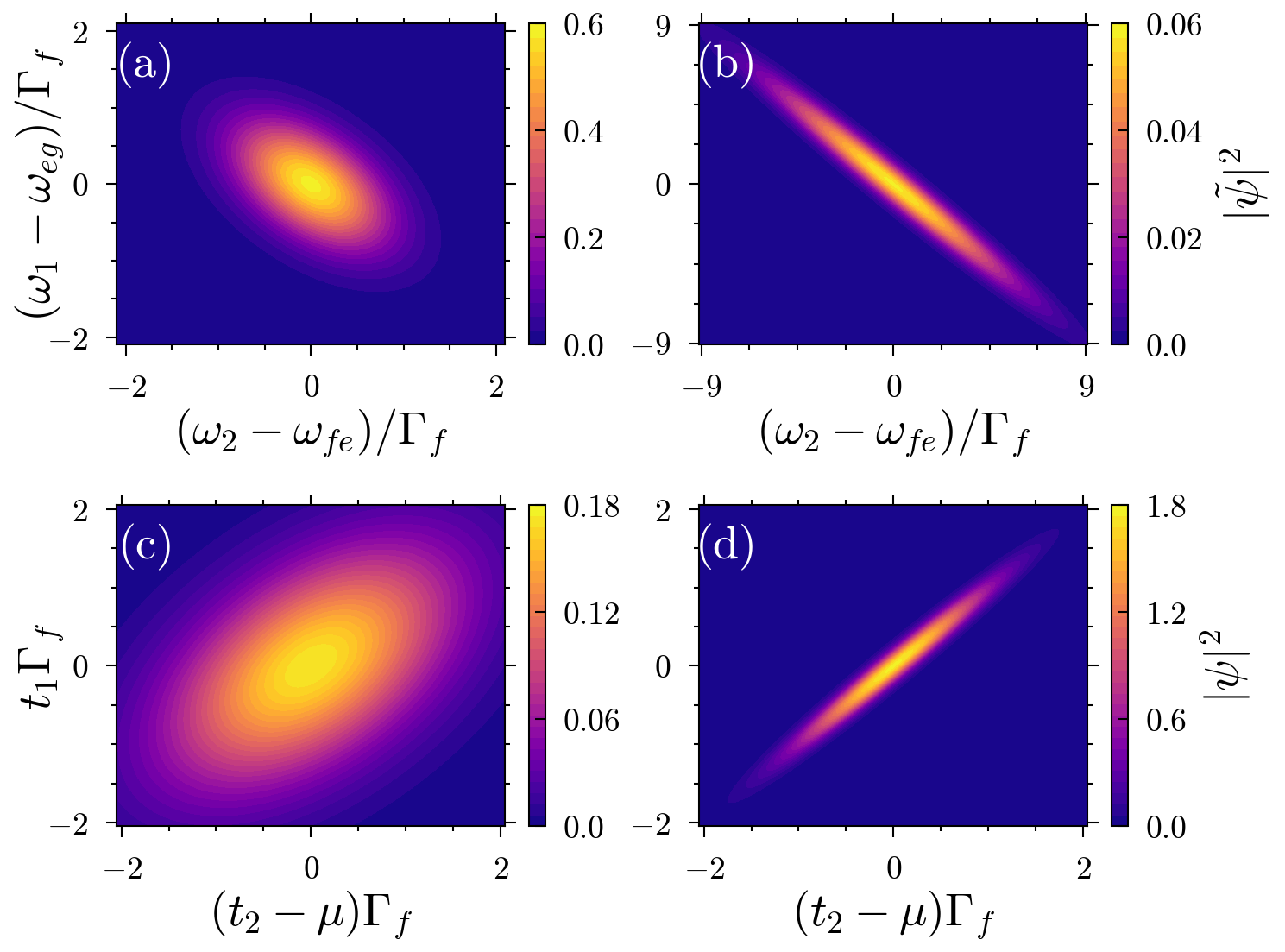}
\caption{Probability density functions in the frequency and time domains of the two-photon state of entangled photons for the ratio: $\Gamma_e/\Gamma_f = 0.5$ ((a), (c)) and $\Gamma_e/\Gamma_f = 5$ ((b), (d)). The values of optimized parameters: (a), (c): $\Omega_{+}=0.79\Gamma_f$, $\Omega_{-}=1.38\Gamma_f$, $\mu=1.62/\Gamma_f$; (b), (d): $\Omega_{+}=1.03\Gamma_f$, $\Omega_{-}=10.82\Gamma_f$, $\mu=0.19/\Gamma_f$.}
\label{fig: entangled_state}
\end{figure}

In Fig.~\ref{fig: entangled_state}, the optimized state (\ref{eq: entanstate}) is illustrated using two-dimensional probability density functions in the frequency and time domains for the $\Gamma_e/\Gamma_f$ ratios of $0.5$ and $5$. Note that the greater the value of the $\Gamma_e/\Gamma_f$ ratio, the more anti-correlated the photons are in terms of the frequency. Their joint spectral distribution becomes increasingly concentrated along the antidiagonal defined by the relation $\omega_1 + \omega_2 = \omega_{eg} + \omega_{fe}$. 
In the regime of $\Gamma_e\gg\Gamma_f$, the distribution in Fig.~\ref{fig: entangled_state}(b) is similar to the one obtained for the optimal state in Fig.~\ref{fig: optimal}(b), reproducing its qualitative character with the dominant peak around the two-photon resonance $\omega_1 + \omega_2 = \omega_{eg} + \omega_{fe}$. Contrary, for $\Gamma_e\ll\Gamma_f$, the qualitative shape of the optimal distribution in Fig.~\ref{fig: optimal}(a) is not reproduced. 
Thus, we conclude that entanglement helps increase the two-photon absorption probability in the virtual regime. Note, however, that the maximal TPA probability achieved with Gaussian shapes reads 0.64. 

\begin{figure}[h]
\includegraphics[width=7cm]{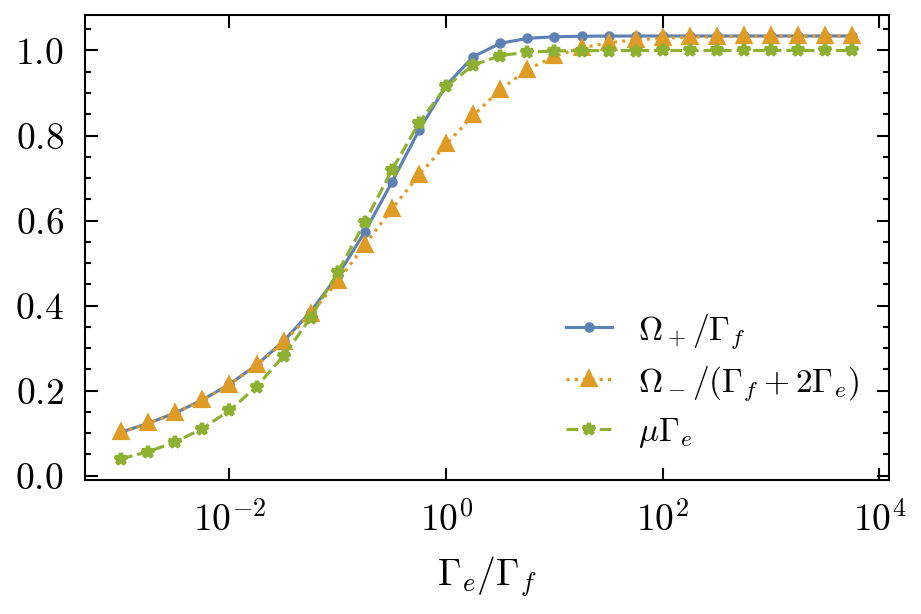}
\caption{ Terms $\Omega_+/\Gamma_f$, $\Omega_-/(\Gamma_f+2\Gamma_e)$, and $\mu\Gamma_e$  for the optimal excitation of the atom by the entangled pair of photons with different mean arrival times as a function of the $\Gamma_e/\Gamma_f$ ratio.}\label{OmegaMu_vs_GammaRatio}
\end{figure}

Details of how the optimal choice of parameters characterizing the pulse depends on the $\Gamma_e/\Gamma_f$ ratio are shown in Fig.~\ref{OmegaMu_vs_GammaRatio}. 
The spectral widths $\Omega_\pm$ are shown in relation to the full widths at half-maxima for the Lorentzian probability distributions of the frequency sum and differences of the optimal light state. 
As the ratio $\Gamma_e/\Gamma_f$ grows, both $\Omega_{-}/(\Gamma_f+2\Gamma_e)$ and $\Omega_+/\Gamma_f$  increase to the limiting value of 1. 
For $\Gamma_e\gg\Gamma_f$, the average time delay of the second photon satisfies the relation $\mu\Gamma_e \approx 1$.
We observe moreover that in this limit $2\sigma^2_{t}\approx \Gamma_f^{-2}$ and  $2\sigma^2_{\omega}\approx \Gamma_e^2$. For comparable $\ket{e}$ and $\ket{f}$ lifetimes we find that the optimized excitation is obtained by the light with the $\Omega_{-}/\Omega_{+}$ ratio being only slightly greater than one. A discussion of the sensitivity of the maximum excitation probability of the atom to the choice of the $\Omega_{+}$ and $\Omega_{-}$ parameters is presented in Appendix \ref{Appendix: sensitivity}.

\section{Results for coherent states}\label{sec:coherent}

\begin{figure}[h]
\begin{subfigure}[b]{0.48\textwidth}
\centering
\includegraphics[width=7cm]{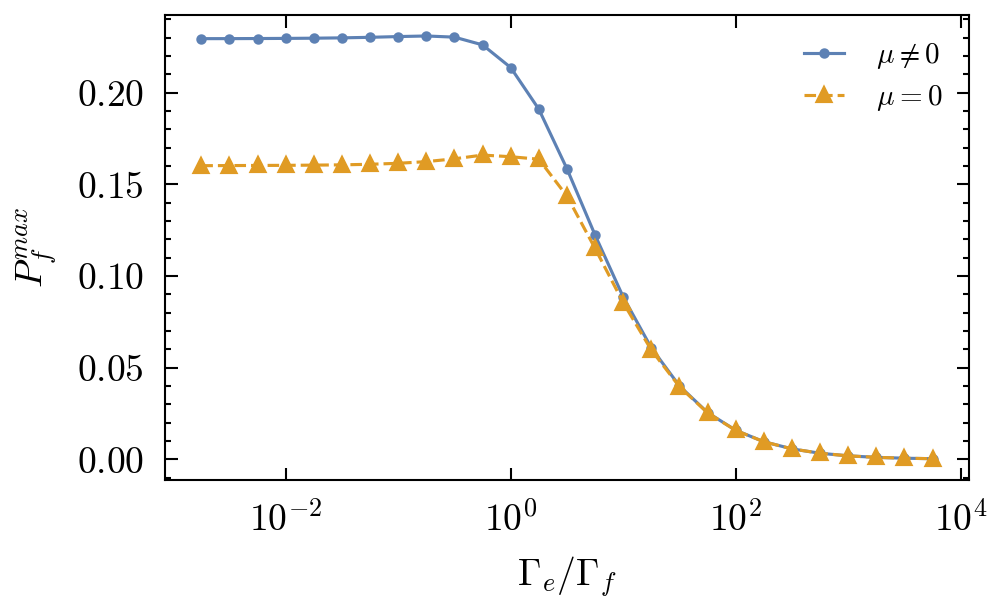}
\end{subfigure} 
\caption{Maximum probability $P^{max}_{f}$ for input coherent states with the mean number of photons $n_1 = n_2 = 1$ and Gaussian shapes as a function of the ratio $\Gamma_{e}/\Gamma_{f}$ with the optimized bandwidth, and with and without optimized time delay $\mu$.}
\label{fig: Coherent pulses for n = 1}
\end{figure}

In this section, we consider the excitation of a three-level atom by two coherent pulses. We assume that the input light is prepared in the pure separable state
\begin{equation}
\ket{\{\alpha_1\}}\otimes \ket{\{\alpha_2\}},
\end{equation}
where
\begin{equation}\label{eq: coherent_state}
\ket{\{\alpha_j\}}= \exp\left(\int_{t_0}^{+\infty}(\alpha_j(t)\hat{a}^{\dagger}_{j}(t)-\alpha_j^{\ast}(t)\hat{a}_{j}(t))dt\right)\ket{0}, 
\end{equation}
is a continuous-mode coherent state \cite{Loudon00} with the amplitude $\alpha_{j}(t)$ such that
\begin{equation}
\int_{t_0}^{+\infty}|\alpha_j(t)|^2dt=n_j,
\end{equation}
where $n_{j}$ is the mean number of photons in the coherent pulse (\ref{eq: coherent_state}). The amplitude of the coherent state can be written in the form $\alpha_{j}(t)=\sqrt{n_j}\alpha_{0j}(t)$, where $\alpha_{0j}(t)\in \mathbb{C}$ and  
\begin{equation}
\int_{t_0}^{+\infty}ds|\alpha_{0j}(s)|^2=1.
\end{equation}
 We assume that the central frequencies of the pulses are close to the transition frequencies of the atom. The evolution of the atom is governed then by the master equation of the form \cite{WisemanMilburn2010,Milburn94}
\begin{equation}\label{eq: master_equation}
\dot{\rho}(t)=-i[\hat{H}+\hat{H}_{drive},\rho(t)]+\sum_{j=1}^{2}\mathcal{D}[\hat{L}_{j}]\rho(t)
\end{equation}
where
\begin{equation}
\hat{H}_{drive}=\sum_{j=1}^{2}i\alpha_j^{\ast}(t)\hat{L}_j-i\alpha_{j}(t)\hat{L}_j^{\dagger},
\end{equation}
and 
\begin{eqnarray}
\mathcal{D}[\hat{L}_{j}]\rho=\hat{L}_j\rho \hat{L}_j^{\dagger}-\frac{1}{2}\{\hat{L}_j^{\dagger}\hat{L}_j,\rho\}
\end{eqnarray}
with $\{\hat{A},\hat{B}\}=\hat{A}\hat{B}+\hat{B}\hat{A}$. The coupling operators are defined by (\ref{eq: coupling_operators}) and the Hamiltonian of the system is given by (\ref{eq: hamiltonian}). By $\rho(t)$ we denote the density operator of the atom at time $t$. The equations for the elements of the density matrix of the atom in the $\ket{g}$, $\ket{e}$, $\ket{f}$ vector basis are given in Appendix \ref{Appendix: master_equation}. As before, we assume that the atom is in the ground state at the initial time. We are interested in the population of the $\ket{f}$ state. For a fair comparison with the results obtained for Fock states in previous sections, we consider photons with Gaussian profiles and assume that 
$n_1=n_2=1$. Figure \ref{fig: Coherent pulses for n = 1} shows the maximum probability value as a function of the $\Gamma_e/\Gamma_f$ ratio. It compares two scenarios: with identical and different average photon arrival times. Optimizing the delay of the second photon brings benefits in the case where the values of $\Gamma_e/\Gamma_f$ are small. When $\Gamma_e\ll\Gamma_f$ the maximum value of the probability of excitation of state $\ket{f}$ approaches $0.23$. 
Note that the maximum value of the excitation probability of a two-level atom by coherent light with one photon on average with a Gaussian profile is $0.48$ \cite{Scarani11}. 
The limiting value of $0.23$ is approximately equal to $(0.48)^2$, which is not a coincidence. Again, we find that in the limit of $\Gamma_e\ll\Gamma_f$, the excitation of the three-level system is optimally realised via two subsequent acts of single-photon absorption.
The value of $0.23$ is achieved for $\Omega_1/\Gamma_e\approx 2.4$, $\Omega_2/\Gamma_f\approx 2.4$, and $\mu\Gamma_e\approx 0.60$. For all the ratios $\Gamma_e/\Gamma_f$ considered, the second photon has a distribution narrower in time (wider in frequency) compared to the first one. In the limit of $\Gamma_e\gg\Gamma_f$, the $\Omega_2/\Omega_1$ ratio is close to $1.3$. 

As the $\Gamma_e/\Gamma_f$ grows, the excitation probability for the coherent states drops due to the lack of entanglement, found to be essential in the regime of $\Gamma_e\gg\Gamma_f$. A comparison with the results for the two-photon state with uncorrelated photons of Gaussian profiles from Sec.~\ref{sec:unentangled} systematically shows that the achieved excitation probability values are lower with coherent states. This can be attributed to the mismatch of photon number statistics with the pair of atomic transitions. 

 In Appendix  \ref{Appendix: Detunings}, the effect of the choice of central pulse frequencies on the efficiency of atomic excitation has been analyzed. We check that for small values of the $\Gamma_e/\Gamma_f$ ratio, it is important to match the central pulse frequencies to both atomic transitions individually. For large values of the $\Gamma_e/\Gamma_f$ ratio, it is possible to ensure efficient atom excitation by entangled photons even if only the two-photon resonance condition is fulfilled. 

\vspace{0.5cm}

\section{Conclusions}\label{sec:conclusions}
In conclusion, we have characterized the role of the light state for TPA in model three-level molecules. It has been shown that the optimization of the state of light must take into account the optical properties of the molecules characterized by their spectral lifetimes $\Gamma_e^{-1}$ and $\Gamma_f^{-1}$. This optimization yields qualitatively different results in regimes defined by the ratio $\Gamma_e/\Gamma_f$. We described the properties of the optimal state leading to the perfect excitation, found in the form corresponding to the time reversal of the two-photon state spontaneously emitted from the molecule. It is a two-photon entangled state whose degree of entanglement increases with the ratio $\Gamma_e/\Gamma_f$. Since the optimal state is not typically realized in laboratory settings, we characterized the excitation probabilities with more feasible profiles of two-photon and coherent states, particularly those described by Gaussian and exponential functions. 
We have shown that the maximum achievable absorption probability is sensitive to the shape of the photons' temporal profiles and the time delay between them.
In the regime of $\Gamma_e \ll \Gamma_f$, we observed that the  TPA process converges to a sequence of two independent single-photon absorptions, occurring at different moments. As a consequence, better excitation efficiency is achieved with uncorrelated photons.
For a growing $\Gamma_e/\Gamma_f$ ratio, we have found an increasing advantage of using entangled photon pairs. The comparison of excitation probabilities obtained for uncorrelated two-photon and coherent states indicates the significant role of photon statistics: it is important for the molecule to interact with exactly one photon per transition. Moreover, for $\Gamma_e \gg \Gamma_f$ the coherent excitation probability is suppressed since the state is not entangled. 

The results obtained in this paper can find application in planning experiments using entangled photon pairs, but clearly the model has limitations and cannot be implemented and discussed in isolation from the physical approximations on which it is based.  
Further work is planned to analyze the optimization of three-level atom excitation when atomic transitions are located closer than the spectral width of the states. 

\section*{Acknowledgements}
The authors are grateful for the financial support by the National Centre for Research and Development, Poland, within the QUANTERA II Programme under Project QUANTERAII/1/21/E2TPA/2023.
All datasets and the numerical codes calculating them are available in the repository \texttt{https://github.com/bojnordsky/TwoPhotonAbsorption}.

\appendix

\section{Proof of the normalization of $P_{f}(t)$}\label{Appendix: Proof}

We would like to prove that $P_{f}(t)\leq 1$ for all ${t\geq t_{0}}$.  For this purpose, referring to the Schmidt decomposition (\ref{eq: sd}) of the state (\ref{eq: tps}) and the triangle inequality, we obtain for the quantity $P_{f}(t)$ the following  upper limit
\begin{equation}
P_{f}(t)\leq \sum_{n=0}^{+\infty} u_{n}^2I_{n},
\end{equation}
where 
\begin{widetext}
 \begin{equation}
I_n=\Gamma_e\Gamma_f \left|\int_{t_0}^{t}dt_2\exp\left[-\left(i\Delta_{2}+\frac{\Gamma_f}{2}\right)(t-t_2)\right]\phi_{n}(t_2)\int_{t_0}^{t_2}dt_1\exp\left[-\frac{\Gamma_e}{2}(t_2-t_1)+i\Delta_1 t_1\right]\xi_{n}(t_1)\right|^2.\end{equation}   \end{widetext}
We prove now that $I_n\leq 1$. In the first step, we show that

\begin{equation}\label{I}
J_{n}=\Gamma_{e} e^{-\Gamma_e t}\left|\int_{t_0}^{t} ds e^{\frac{1}{2}\left(\Gamma_e +2i\Delta_1\right)s}\xi_{n}(s)\right|^2\leq 1.
\end{equation}
Using the Cauchy-Schwarz inequality, we obtain 
\begin{equation}
J_{n}\leq \Gamma_{e} e^{-\Gamma_e t}\int_{t_0}^{t}ds\, e^{\Gamma_e s} \int_{t_0}^{t}ds\left|\xi_{n}(s)\right|^2.
\end{equation}
Then referring to the normalization condition for the amplitude $\xi_{n}$, we get 
\begin{equation}
J_{n}\leq \left(1-e^{-\Gamma_{e}(t-t_0)}\right)\leq 1. 
\end{equation}
Now let us return to the inequality for $I_{n}$. By the Cauchy-Schwarz inequality, we can get the following upper limit of $I_n$
\begin{widetext}
\begin{equation}
I_n\leq \Gamma_e\Gamma_f e^{-\Gamma_{f}t} \int_{t_0}^{t}ds |\phi_{n}(s)|^2 
\int_{t_0}^{t}dt_2 e^{\left(\Gamma_f-\Gamma_e\right) t_2}\left|\int_{t_0}^{t_2}dt_1e^{\frac{1}{2}\left(\Gamma_e +2i\Delta_1\right)t_1} \xi_{n}(t_1)\right|^2.\end{equation}
\end{widetext}
Using the normalization condition for $\phi$ and the inequality (\ref{I}), we have
\begin{equation}\label{norm_In}
I_n\leq 1- e^{-\Gamma_{f}(t-t_0)}\leq 1.
\end{equation}
Finally, by the inequalities (\ref{norm_In}) and (\ref{eq: norm_u}), we obtain that $P_{f}(t)\leq 1$ for any $t\geq t_{0}$. 

\section{$P_{f}(t)$ as the inner product}\label{Appendix: Inner_Product}

The formula for $P_{f}(t^{\star})$ can be treated as the module squared of the inner product of the two functions from the space $L^{2}\left([t_0,t^{\star}]\otimes[t_0,t^{\star}] \right)$. Here, we restrict ourselves to the case of resonances and take $t_0\to -\infty$, but of course, similar things can be done in an extended version beyond resonances and for a finite $t_0$. Let us notice that then we can rewrite (\ref{eq: Pf}) in the form 
\begin{equation} \label{eq: Pf_inner product}
P_{f}(t^{\star})= \left|\int_{-\infty}^{t^{\star}}dt_2\int_{-\infty}^{t^{\star}}dt_1\Psi_{0}(t_2,t_1)\Psi(t_2,t_1)\right|^2,
\end{equation}
where
\begin{equation}\label{eq: optimal distribution}
 \Psi_{0}(t_2,t_1)=\Gamma_e\Gamma_f e^{-\Gamma_f t^{\star}}e^{\frac{1}{2}(\Gamma_f-\Gamma_e)t_2+\frac{1}{2}\Gamma_{e}t_{1} }\chi_{-\infty<t_1<t_2<t^{\star}}.  
\end{equation}
Note that we deal with two normalized functions. Referring to this, it is easy to determine the maximal value of $P_{f}(t^{\star})$. The maximal value of a module of an inner product of normalized functions is $1$ and it is obtained when the functions are proportional. Hence we obtain the formula for the optimal state 
(\ref{eq: optstate1})
with the optimal temporal distribution (\ref{eq: optimal distribution}).

\section{Sensitivity}\label{Appendix: sensitivity}

\begin{figure}[h]
		\includegraphics[width=8.5cm]{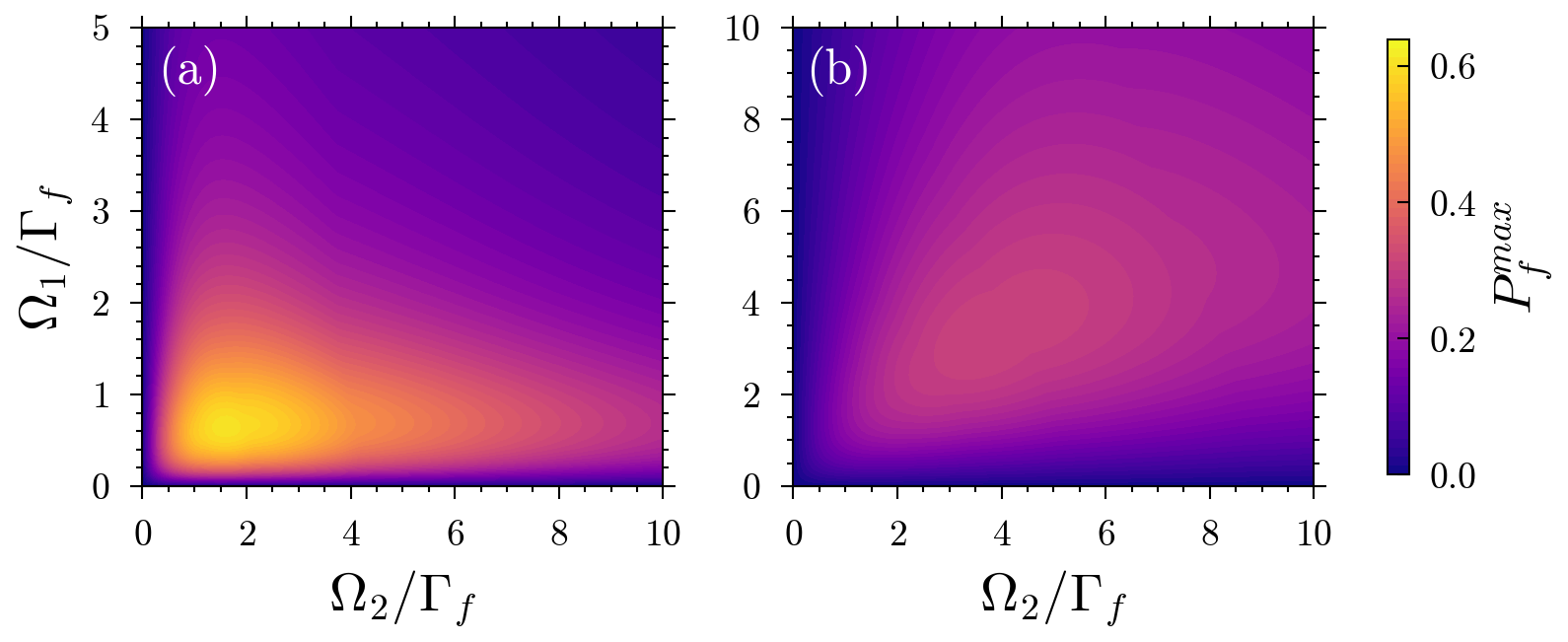}
\caption{Maximum probability $P_f^{max}$ as a function of $\Omega_1/\Gamma_f$ and
$\Omega_{2}/\Gamma_f$ for the input two-photon state of unentangled photons with the Gaussian shapes with the optimized time delay $\mu$ for the ratio $\Gamma_e/\Gamma_f$ equal to $0.5$ (a) and $5$ (b). 
}
 \label{fig:Uncorr_heatMap}
\end{figure}

The sensitivity of the maximal probability value with respect to the spectral widths of the two unentangled photons for the case with the optimized time delay is shown in Fig.~\ref{fig:Uncorr_heatMap}. 
Here, we discuss the results for the photons of the Gaussian shapes. We observe the dependence of the final state probability to be more symmetric in the case of greater $\Gamma_e/\Gamma_f$, as expected from the analysis of the optimal state.
In the regime of small $\Gamma_e/\Gamma_f$ ratio, the second photon needs to be broader in frequencies, i.e., narrower in time to match the finite lifetime of the $\ket{e}$ state. At the same time, the calculated probability of the excitation of state $\ket{f}$ is less sensitive with respect to the broadening $\Omega_2$ of the second photon.

\begin{figure}[th]
		\includegraphics[width=8.5cm]{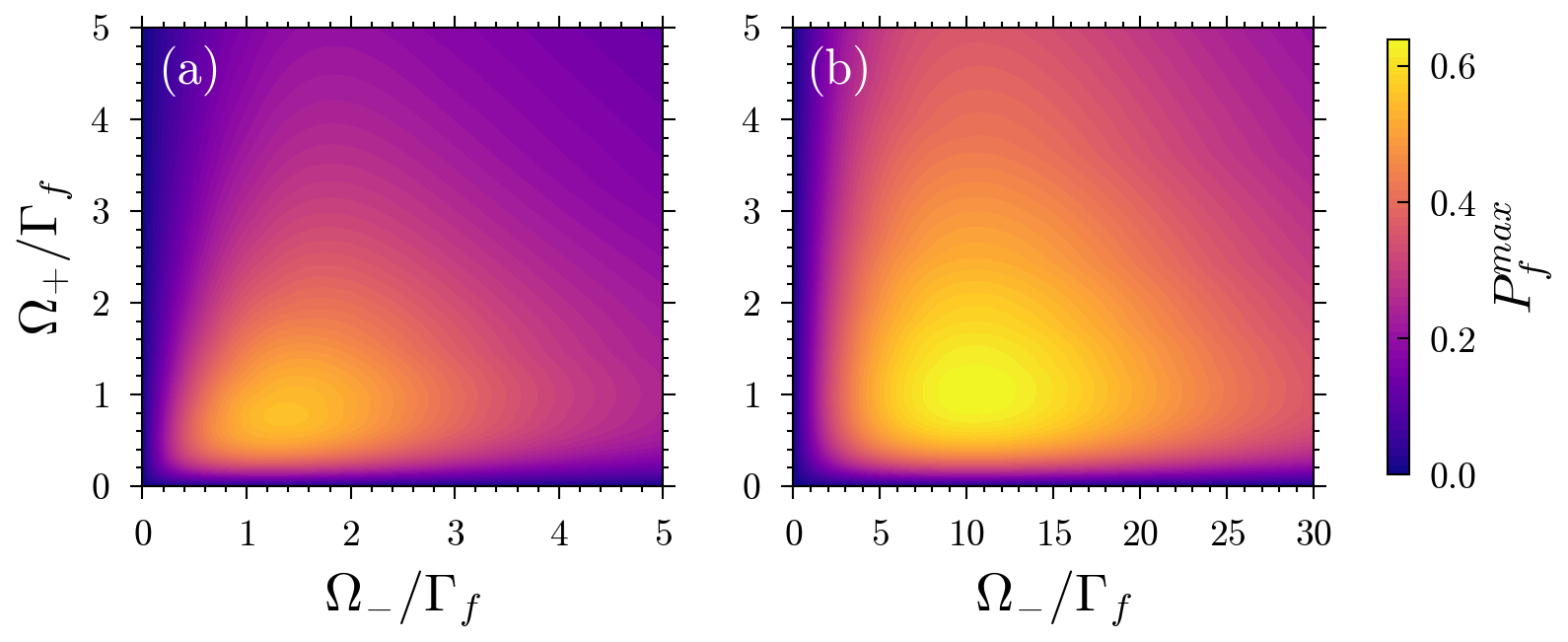}
\caption{Maximum probability $P_{f}^{max}$ as a function of $\Omega_{+}/\Gamma_f$ and
$\Omega_{-}/\Gamma_{f}$ for the input two-photon state of entangled photons with the optimized time delay $\mu$ for the ratio $\Gamma_e/\Gamma_f$ equal to $0.5$ (a) and $5$ (b).} 
 \label{fig: corr_heatMap}
\end{figure}

The sensitivity plot of the maximal probability value to the parameters characterizing the state (\ref{eq: entanstate}) of entangled photons is shown in Fig.~\ref{fig: corr_heatMap} for $\Gamma_e/\Gamma_f = 0.5$ and $5$. For the small $\Gamma_e/\Gamma_f$ ratio, due to the relatively low degree of entanglement, the sensitivity of the achieved excitation probability to variations in $\Omega_+$ and $\Omega_-$ is similar. For large $\Gamma_e/\Gamma_f$ ratio, we observe an asymmetry: it is important for $\Omega_-$ to be large in order to achieve a high degree of entanglement, but the sensitivity to the exact value of $\Omega_-$ is limited. Contrary, sensitivity to the value of $\Omega_+$ is greater, since the total energy of the photon pair needs to match $\omega_{fg}$.

\section{Results for two-photon states of unentangled photons. Other profiles}\label{Appendix: Rising_Decaying_Shapes}

\subsection{Rising exponential pulses}

Let us consider the case when the atom is driven by the light in the two-photon state for the unentangled photons with the temporal profiles: 
\begin{eqnarray}\label{eq: rising1}
\xi(t)= \left\{ \begin{array}{lll}
\sqrt{\Omega}_{1}e^{\frac{\Omega_1}{2}t}& \mathrm{for}& t\leq 0\\
0 & \mathrm{for}& t>0
\end{array}\right.,
\end{eqnarray}
\begin{equation}\label{eq: rising2}
\phi(t)= \left\{ \begin{array}{lll}
\sqrt{\Omega}_{2}e^{\frac{\Omega_2}{2}t}& \mathrm{for}& t\leq 0\\
0 & \mathrm{for}& t>0
\end{array}\right.,
\end{equation}
where $\Omega_{i}, i=1,2$ are positive parameters. 
In this case the probability of the finite-state excitation (\ref{eq: Pf}) is given by the analytical formulas:
\begin{equation}
P_{f}(t) = \frac{16\Gamma_{e}\Gamma_{f}\Omega_1\Omega_2 e^{(\Omega_1+\Omega_2)t}}{(4\Delta_1^2+(\Omega_{1}+\Gamma_{e})^2)(4(\Delta_1+\Delta_2)^2+(\Omega_1+\Omega_2+\Gamma_f)^2)}
\end{equation}
for time $t\leq 0$  and $P_{f}(t)=P_{f}(0)e^{-\Gamma_{f}t}$ for $t>0$. 
One can verify that then the probability $P_{f}(t)$ reaches its maximum value of 
\begin{equation}
P_{f}^{max}= \frac{64 \frac{\Gamma_e}{\Gamma_f} \left(\sqrt{1+8\frac{\Gamma_e}{\Gamma_f}}-1\right)}{\left(\frac{4\Gamma_e}{\Gamma_f}+\sqrt{1+\frac{8\Gamma_e}{\Gamma_f}}-1\right)^2\left(3+\sqrt{1+\frac{8\Gamma_e}{\Gamma_f}}\right)} ,
\end{equation}
at time $t=0$ for the parameters 
\begin{equation}\label{omegas}
\Omega_{1}= \frac{\sqrt{\Gamma_{f}^2+8\Gamma_{e}\Gamma_{f}}-\Gamma_{f}}{4}, \; \Omega_{2}=\Omega_1+\Gamma_{f}, \Delta_{i}=0.  
\end{equation}

The dependence of $P^{max}_f$ on the $\Gamma_e/\Gamma_f$ ratio is depicted in Fig. \ref{fig: expon_pulses}. It is seen that the effectiveness of the excitation for the rising exponential pulses depends on the $\Gamma_e/\Gamma_f$ ratio. One can check that the probability $P^{max}_{f}$ approaches the value of one
when $\Gamma_e\ll \Gamma_f$ and it goes to zero when $\Gamma_e\gg\Gamma_f$. 

Clearly, for the rising exponential pulses, defined by (\ref{eq: rising1}) and (\ref{eq: rising2}),  the mean times of the photon arrivals are given respectively by $-1/\Omega_1$ and $-1/\Omega_2$. Thus the difference between them for parameters given by (\ref{omegas}) is 
\begin{equation}
\mu = \frac{16}{\Gamma_f\left(\sqrt{1+\frac{8\Gamma_e}{\Gamma_f}}-1\right)\left(\sqrt{1+\frac{8\Gamma_e}{\Gamma_f}}+3\right)}. 
\end{equation}
One can prove that if $\Gamma_e/\Gamma_f \to \infty$, then  $\mu\Gamma_f \to 0$ and if $\Gamma_e/\Gamma_f\to 0$, then $\mu \Gamma_f\to \infty$.

\subsection{Decaying exponential pulses}

\begin{figure}[h]
		\includegraphics[width=7cm]{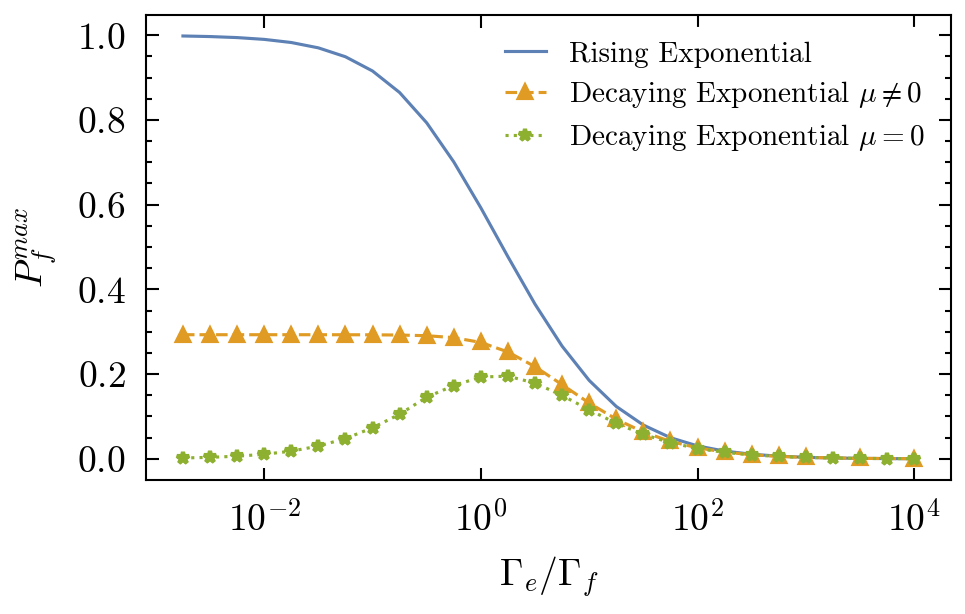}
	        \caption{Maximum probability $P^{max}_{f}$ for the input two-photon states of unentangled photons with the rising and decaying exponential shapes as a function of the ratio $\Gamma_{e}/\Gamma_{f}$ with the optimized bandwidth and with and without optimized time shift $t_s$. }
   \label{fig: expon_pulses}
\end{figure}

Let us consider the two-photon state for the unentangled photons with the temporal profiles: 
\begin{eqnarray}\label{eq: decaying1}
\xi(t)= \left\{ \begin{array}{lll}
0 & \mathrm{for}& t<0\\
\sqrt{\Omega}_{1}e^{-\frac{\Omega_1}{2}t}& \mathrm{for}& t\geq 0
\end{array}\right.,
\end{eqnarray}
\begin{equation}\label{eq: decaying2}
\phi(t)= \left\{ \begin{array}{lll}
0 & \mathrm{for}& t<t_s\\
\sqrt{\Omega}_{2}e^{-\frac{\Omega_2}{2}\left(t-t_s\right)}& \mathrm{for}& t\geq t_s
\end{array}\right.,
\end{equation}
where $\Omega_{1},\Omega_2>0$, and $t_s$ is a shift time of the second pulse. We consider two scenarios, one with a fixed $t_s=0$ and the other with $t_s$ optimized to achieve the maximum value of the probability of finite-state excitation. For the given two-photon state, we obtain 
\begin{equation}
P_{f}(t)=\frac{\Gamma_e\Gamma_f\Omega_1\Omega_2 e^{-\Gamma_ft+\Omega_2t_s}}{|abc|^2}\left|b(e^{ct}-e^{ct_s})-c(e^{bt}-e^{bt_s})\right|^2,
\end{equation}
for $t\geq t_s$, where
\begin{align}
a &= i\Delta_1+\frac{1}{2}(\Gamma_e-\Omega_1)\\
b& = i\Delta_2+\frac{1}{2}(\Gamma_f-\Gamma_e-\Omega_2)\\
c& = i(\Delta_1+\Delta_2)+\frac{1}{2}(\Gamma_f-\Omega_1-\Omega_2)
\end{align}
and zero in other cases. The results of the numerical optimization of the finite-state excitation are shown in Fig.~\ref{fig: expon_pulses}. In the area where the time shift benefits, $t_s$ has an approximate value of $1/\Omega_2+1/\Gamma_e-1/\Omega_1$. In the limiting case, when $\Gamma_e \ll \Gamma_f$, the maximum value of the probability of excitation of the final state reaches $0.29$, which corresponds to the square of the maximum probability of excitation of a two-level atom driven by a single-photon pulse with a decaying exponential shape, $(0.54)^2$.

\section{Set of equations for coherent excitation}\label{Appendix: master_equation}

Working in the rotation frame (\ref{eq: rotating_frame}), without losing the generality of considerations, we can assume that $\alpha_{j}(t)$ is a function with real values. 
Then from Eq. (\ref{eq: master_equation}), we obtain the set of differential equations:
\begin{widetext}
\begin{align}
\dot{\rho}^{ff}(t)&= -\sqrt{\Gamma_f n_2}\alpha_{02}(t)\left(\rho^{ef}(t)+\rho^{fe}(t)\right)-\Gamma_{f}\rho^{ff}(t)\\
\dot{\rho}^{ef}(t)&= i\Delta_{2}\rho^{ef}(t)-\sqrt{\Gamma_e n_1}\alpha_{01}(t)\rho^{gf}(t)+\sqrt{\Gamma_f n_2}\alpha_{02}(t)\left(\rho^{ff}(t)-\rho^{ee}(t)\right)-\frac{1}{2}\left(\Gamma_{e}+\Gamma_{f}\right)\rho^{ef}(t)\\
\dot{\rho}^{gf}(t)&= i(\Delta_1+\Delta_{2})\rho^{gf}(t)+\sqrt{\Gamma_e n_1}\alpha_{01}(t)\rho^{ef}(t)-\sqrt{\Gamma_f n_2}\alpha_{02}(t)\rho^{ge}(t)-\frac{\Gamma_{f}}{2}\rho^{gf}(t)\\
\dot{\rho}^{ee}(t)&=-\sqrt{\Gamma_e n_1}\alpha_{01}(t)(\rho^{ge}(t)+\rho^{eg}(t))+\sqrt{\Gamma_f n_2}\alpha_{02}(t)(\rho^{fe}(t)+\rho^{ef}(t))-\Gamma_{e}\rho^{ee}(t)+\Gamma_{f}\rho^{ff}(t)\\
\dot{\rho}^{ge}(t)&=i\Delta_{1}\rho^{ge}(t)+\sqrt{\Gamma_f n_2}\alpha_{02}(t)\rho^{gf}(t)+\sqrt{\Gamma_e n_1}\alpha_{01}(t)\left(\rho^{ee}(t)-\rho^{gg}(t)\right)-\frac{\Gamma_{e}}{2}\rho^{ge}(t)\\
\dot{\rho}^{gg}(t)&= \sqrt{\Gamma_e n_1}\alpha_{01}(t)\left(\rho^{eg}(t)+\rho^{ge}(t)\right)+\Gamma_{e}\rho^{ee}(t),
\end{align}
\end{widetext}
where ${\rho}^{ef}(t)=({\rho}^{fe}(t))^{\dagger}$ and ${\rho}^{ge}(t)=({\rho}^{eg}(t))^{\dagger}$. We assume that initially, the atom is in the ground state, so we consider the following initial conditions: $\rho^{gg}(t_0)=1$ and all the other elements are equal to zero at $t=t_0$. 	

\section{Influence of the detunings}\label{Appendix: Detunings}

\begin{figure}
\centering
\includegraphics[width=8.5cm]{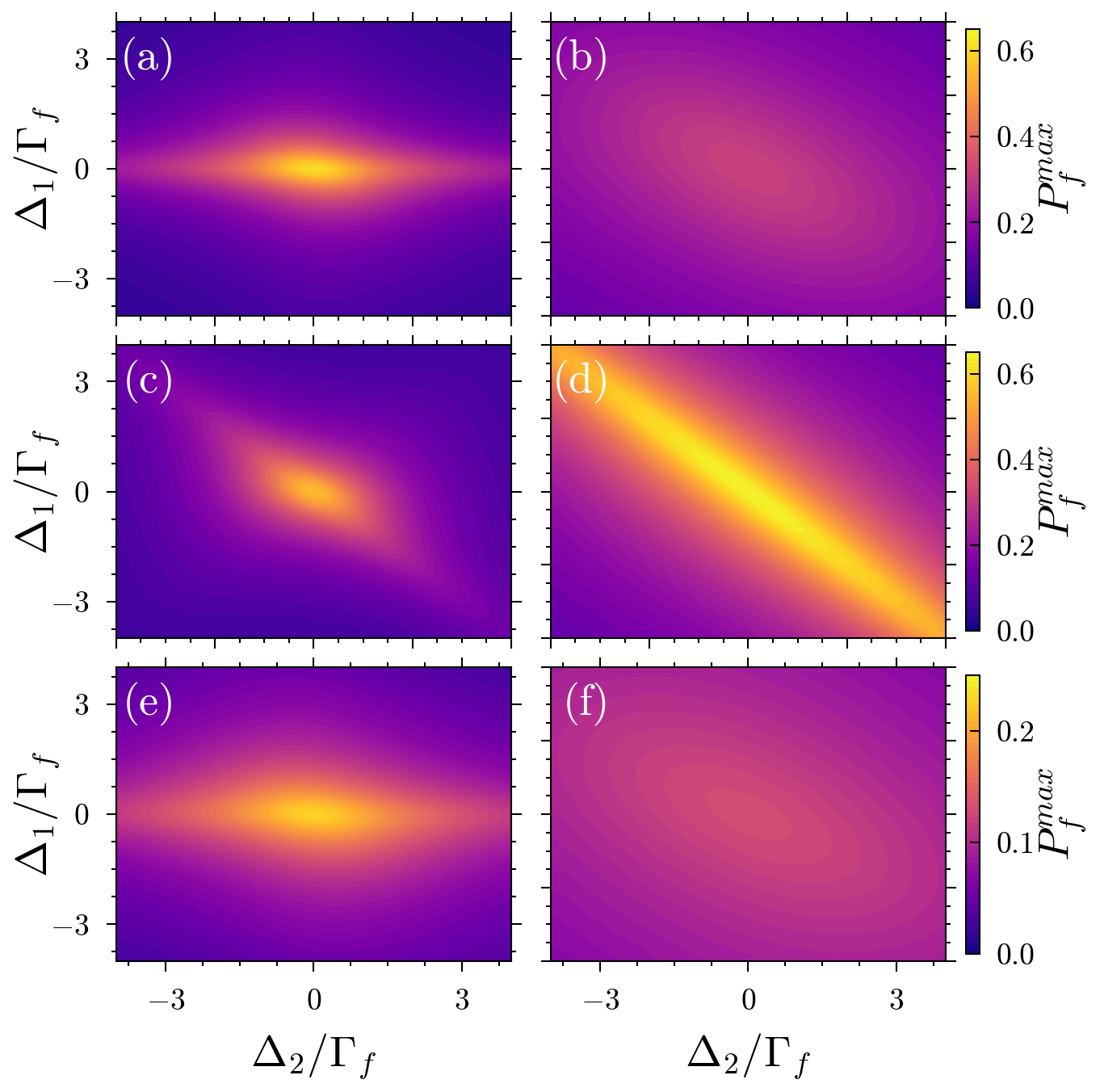}
\caption{Maximum probability $P^{max}_{f}$ as a function of $\Delta_1/\Gamma_f$ and $\Delta_2/\Gamma_f$
for the $\Gamma_e/\Gamma_f$ ratio equal to $0.5$ (a, c, e) and $5$ (b,d,f) for the two-photon state with unentangled (a,b) and entangled photons (c,d), and the coherent states  (e,f). In all cases, photons with Gaussian profiles were considered. }
 \label{fig: detunings}
\end{figure}

Figure \ref{fig: detunings} shows the effect of optimizing the parameters of the different states of light for fixed values of detunings $\Delta_1$ and $\Delta_2$. It means that we determine the maximum probability values $P_{f}(t)$ that can be achieved for the central frequencies different from the frequency of transitions in the atom. We study this problem for the two-photon states with entangled and unentangled photons and for the coherent states. We consider photons with Gaussian profiles. We compare here two cases with the $\Gamma_e/\Gamma_f$ ratio of $0.5$ and $5$. For the $\Gamma_e/\Gamma_f$ ratio of $0.5$, for each of the analyzed states, we observe that achieving effective atomic excitation strongly depends on matching both central frequencies to the atomic transition values. A qualitatively different behaviour of the system is observed for the case when the $\Gamma_e/\Gamma_f$  ratio is equal to $5$. In this case, it becomes crucial to fulfil the condition of a double resonance, i.e. $\omega_{01}+\omega_{02}=\omega_{fg}$. This is particularly strongly visible for the entangled photon pair. In this case, the departure from single resonances while satisfying the double resonance condition over a wide range of detunings does not impede the efficient excitation of the atom.

 \end{document}